\documentclass{article}

    \usepackage[preprint,nonatbib]{neurips_data_2021}
    \usepackage[numbers]{natbib}

\usepackage[utf8]{inputenc} 
\usepackage[T1]{fontenc}    
\usepackage[hyphens]{url}
\usepackage[breaklinks=true]{hyperref}
\usepackage{booktabs}       
\usepackage{amsfonts}       
\usepackage{nicefrac}       
\usepackage{microtype}      
\usepackage{multirow}
\usepackage[utf8]{inputenc}
\usepackage{algorithmic}
\usepackage{graphicx}
\graphicspath{{./}}
\usepackage{textcomp}
\usepackage{xcolor}
\usepackage{enumitem}
\usepackage[numbib]{tocbibind}
\usepackage{cleveref}
\usepackage{subfig}
\usepackage{amssymb}
\usepackage{pifont}

\newcommand{\code}[1]{\texttt{#1}}

\usepackage{pgfplots, pgfplotstable}
\pgfplotsset{compat=1.16}

\pdfoutput=1

\begin{document}

\title{CodeNet:
A Large-Scale AI for Code Dataset for Learning a Diversity of Coding Tasks}
\newcommand*{\affaddr}[1]{#1} 
\newcommand*{\affmark}[1][*]{\textsuperscript{#1}}

\author{
\textbf{Ruchir Puri}\affmark[1],
\textbf{David S. Kung}\affmark[1],
\textbf{Geert Janssen}\affmark[1],
\textbf{Wei Zhang}\affmark[1], \\ 
\textbf{Giacomo Domeniconi}\affmark[1],
\textbf{Vladimir Zolotov}\affmark[1],
\textbf{Julian Dolby}\affmark[1],
\textbf{Jie Chen}\affmark[2,1], \\
\textbf{Mihir Choudhury}\affmark[1],
\textbf{Lindsey Decker}\affmark[1],
\textbf{Veronika Thost}\affmark[2,1],
\textbf{Luca Buratti}\affmark[1], \\
\textbf{Saurabh Pujar}\affmark[1],
\textbf{Shyam Ramji}\affmark[1],
\textbf{Ulrich Finkler}\affmark[1],
\textbf{Susan Malaika}\affmark[3],
\textbf{Frederick Reiss}\affmark[1]
\\
\\
\affaddr{\affmark[1]IBM Research} \\
\affaddr{\affmark[2]MIT-IBM Watson AI Lab} \\ 
\affaddr{\affmark[3]IBM Worldwide Ecosystems}
}

\maketitle
\begin{abstract}
Over the last several decades, software has been woven into the fabric of every aspect of our society. As software development surges and code infrastructure of enterprise applications ages, it is now more critical than ever to increase software development productivity and modernize legacy applications. Advances in deep learning and machine learning algorithms have enabled breakthroughs in computer vision, speech recognition, natural language processing and beyond, motivating researchers to leverage AI techniques to improve software development efficiency. Thus, the fast-emerging research area of “AI for Code” has garnered new interest and gathered momentum. 
In this paper, we present a large-scale dataset \textit{CodeNet}, consisting of over 14 million code samples and about 500 million lines of code in 55 different programming languages, which is aimed at teaching AI to code. In addition to its large scale, CodeNet has a rich set of high-quality annotations to benchmark and help accelerate research in AI techniques for a variety of critical coding tasks, including code similarity and classification, code translation between a large variety of programming languages, and code performance (runtime and memory) improvement techniques. Additionally, CodeNet provides sample input and output test sets for 98.5\% of the code samples, which can be used as an oracle for determining code correctness and potentially guide reinforcement learning for code quality improvements. As a usability feature, we provide several pre-processing tools in CodeNet to transform source code into representations that can be readily used as inputs into machine learning models. Results of code classification and code similarity experiments using the CodeNet dataset are provided as a reference. We hope that the scale, diversity and rich, high-quality annotations of CodeNet will offer unprecedented research opportunities at the intersection of AI and Software Engineering.

\end{abstract}

\section{Introduction}
\label{intro}
\noindent
There is a growing trend towards leveraging AI for building tools that support software engineering and development~\cite{allamanis2018survey, yang2020survey}. AI can manipulate and generate computer code, but can it do so with high quality? Many researchers are fascinated by this possibility, encouraged by AI successes in other domains and tantalized by the vision of computers programming computers. Some recent deep-learning models \cite{chen2021evaluating,Lachaux2020} for code have received a lot of publicity: trained on vast amounts of data and using novel architectures with billions of parameters, they sometimes generate surprisingly plausible code.

Given the success of non-AI tools for code, why should we consider AI to augment or possibly replace them?  Firstly, AI can help refine and re-tune the heuristics used by traditional coding tools. Secondly, based on the training data from past experience, AI can help prioritize when there is more than one sound answer \cite{wang2018machine}. Thirdly, an AI-based tool may handle incomplete or invalid code more robustly, thus expanding its scope. Finally, AI can incorporate signals usually ignored by traditional tools for code, such as the natural language in identifiers or comments.

In the enterprise environment, developers often face code written by large teams over many years and geographies. Developers must manipulate such code to modernize it, fix bugs, improve its performance, evolve it when requirements change, make it more secure, and/or comply with regulations. 
These tasks are challenging, and it is crucial to provide tool support for developers to be more productive at performing them. 
It is well known that the latest advancements in deep learning algorithms rely on best-of-breed datasets, such as ImageNet, to create increasingly complex and powerful models. 
In this paper, we present "CodeNet", a first-of-its-kind dataset in scale, diversity, and quality, to accelerate the algorithmic advances in AI for Code. 

To promote widespread adoption of CodeNet, we will be launching contests involving use cases based on the dataset. The first contest ~\cite{tier1contest} will focus on diversity, inclusion and spurring interest among aspiring data scientists. We are partnering with the Global Women in Data Science organization (with presence in over 50 countries) founded by Stanford University \cite{wids} and targeting teams with at least fifty percent women. 
We are planning follow-up contests that target experienced AI practitioners. 

The rest of the paper is organized as follows. Section 2 introduces the CodeNet dataset. Related datasets are discussed in Section 3, and the differentiation of CodeNet with respect to these related datasets is elaborated in Section 4. Section 5 describes how CodeNet was curated and Section 6 enumerates the usability features of CodeNet with several pre-processing tools to transform source codes into representations that can be readily used as inputs into machine learning models. Section 7 discusses the upcoming CodeNet contest and Section 8 describes important baseline experiments with the CodeNet dataset. Section 9 presents further uses of the CodeNet dataset and Section 10 concludes the paper.
\section{The CodeNet Dataset}
\label{overview}
The CodeNet dataset consists of a large collection of code samples
with extensive metadata. It also contains documented tools to
transform code samples into intermediate representations and to access 
the dataset and make tailored selections.  Our goal is to provide the
community with a large, high-quality curated dataset that can be used to
advance AI techniques for source code.

CodeNet is derived from the data available on two online judge websites: AIZU~\cite{aizu} and AtCoder~\cite{atcoder}.  Online judge
websites pose programming problems in the form of courses and
contests. The dataset consists of submissions to these problems, which are
judged by an automated review process for correctness.  Problem
descriptions, submission outcomes, and associated metadata are
available via various REST APIs.

\begin{figure}
\centering
  \subfloat[Languages]{
    {\includegraphics[height=5cm]{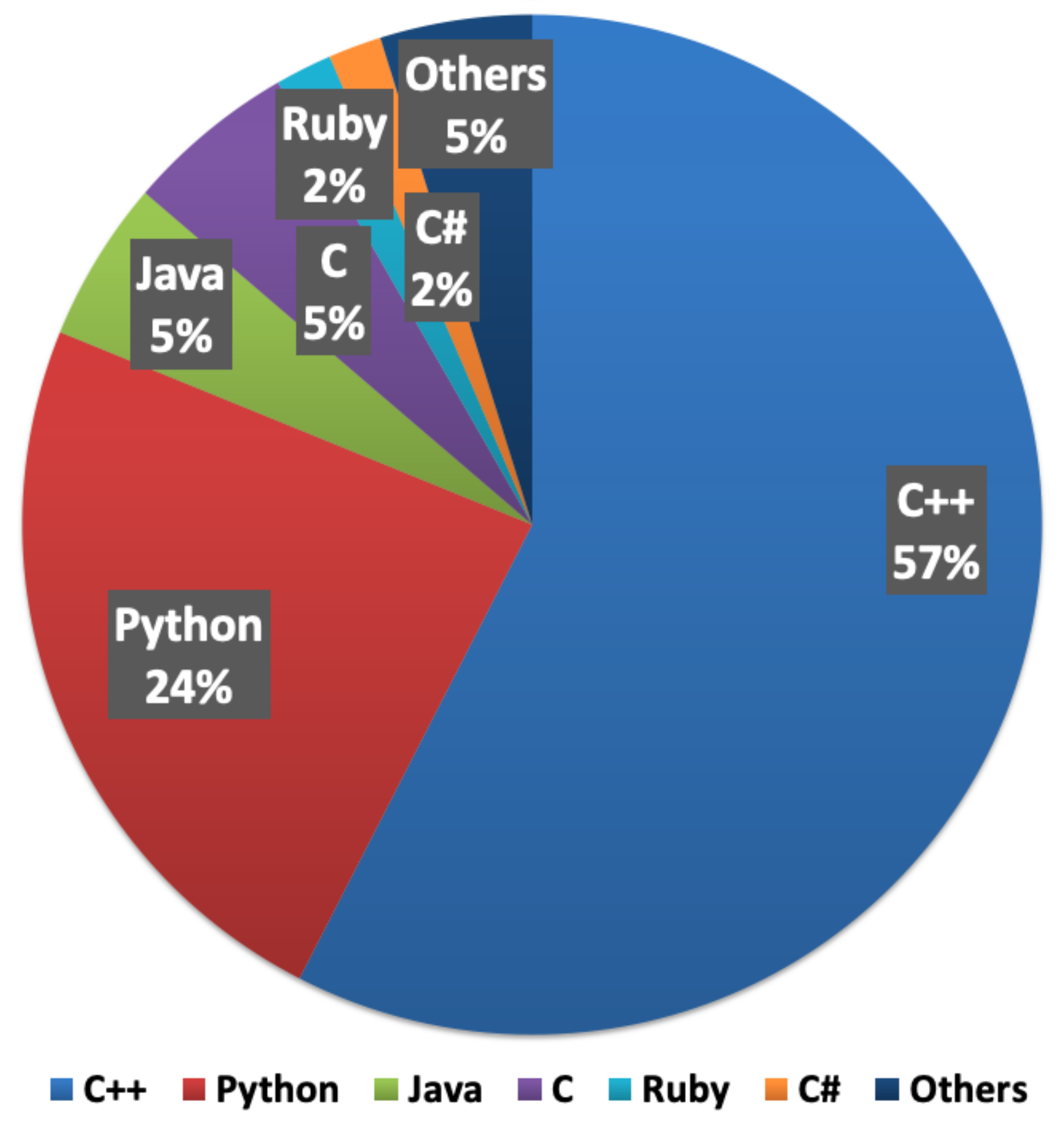}   
    }\label{fig:lang}
  }
  \subfloat[Status]{
    {\includegraphics[height=5cm]{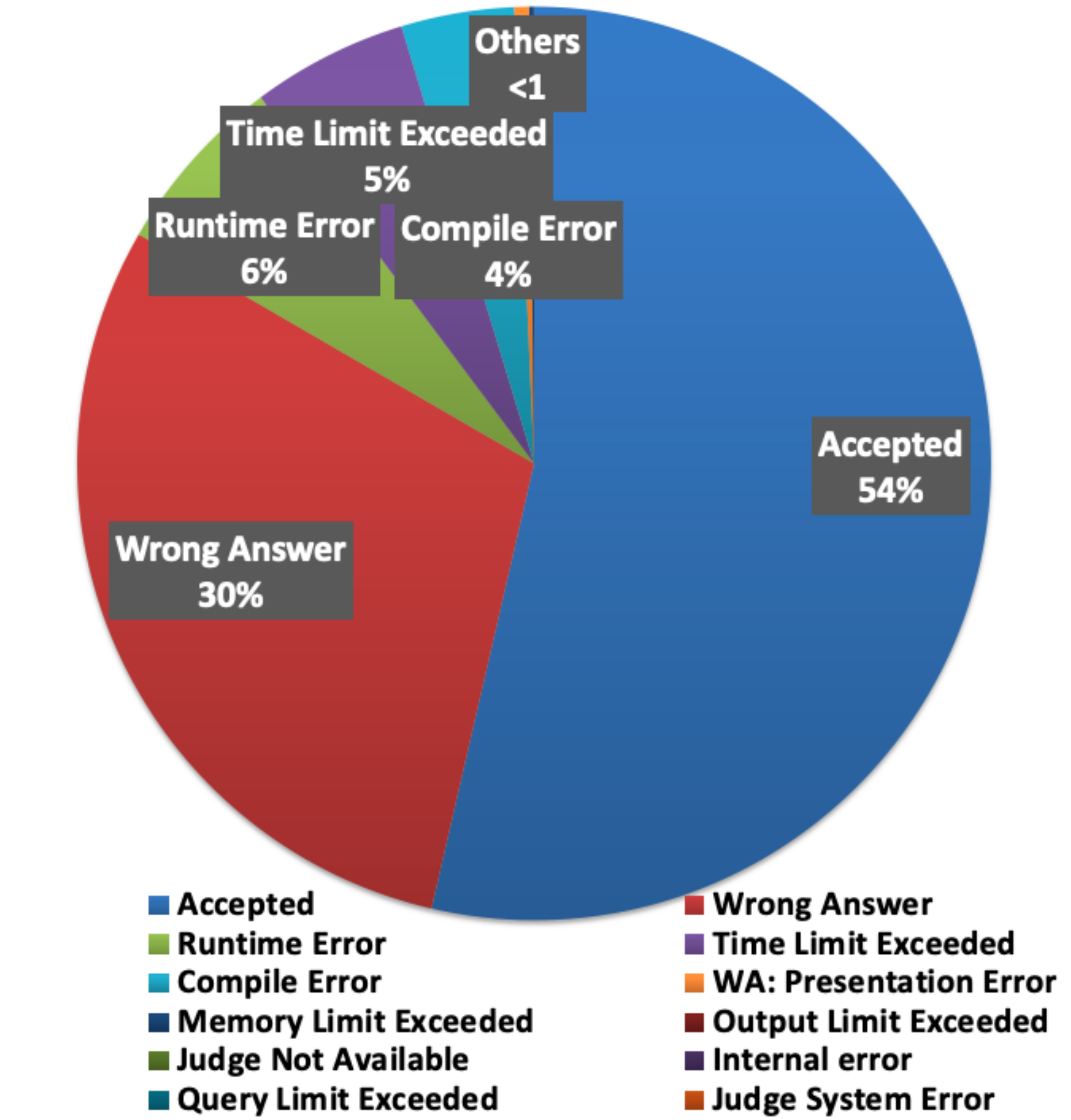}   
    }\label{fig:status}
  }
  \caption{Percentage of submissions per language (left) and per status (right).}
  \label{fig:lang-status}
\end{figure}

\textbf{Scale and Statistics.}
CodeNet contains a total of 13,916,868 submissions, divided into
4053 problems. Among the
submissions, 53.6\% (7,460,588) are accepted (compilable and pass the prescribed tests), 29.5\% are marked with
wrong answer, and the remaining rejected due to their failure to meet
run time or memory requirements. To our knowledge, this is the largest dataset so far among
similar kinds. Submissions are in
55 different languages; 95\% of them are coded in C++, Python, Java,
C, Ruby, and C\#. C++ is the most common language, with 8,008,527
submissions (57\% of the total), of which 4,353,049 are accepted. 
With
the abundance of code samples, users can extract large benchmark
datasets that are customized to their downstream use.
See
Figure~\ref{fig:lang-status} for a summary.

\textbf{Diversity.}
The problems in CodeNet are mainly pedagogical and range from elementary exercises to sophisticated problems that require
advanced algorithms. The submitters range from beginners to experienced coders. Some submissions are correct while others contain
different types of errors, accordingly labeled. The submissions are
in many different languages.

\textbf{Code Samples.}
Each code sample is a single file and includes
inputting the test cases and printing out the computed results. The
file name uses standard extensions that denote the programming
language, e.g., \code{.py} for Python.  The
majority of code samples contain only one function, although submissions to more complex
problems might have several functions.

\textbf{Metadata.}
The metadata enables data queries and selections among the large
collection of problems, languages, and source files.  The metadata is
organized in a two level hierarchy. The first is the dataset
level, which describes all problems. The second is the problem level,
which details all the submissions to a single
problem. Metadata and data are separated in the dataset structure.

At the dataset level, a single CSV file lists all problems and their
origins, along with the CPU time and memory limits set for them.
Additionally, every problem has an HTML file with a detailed
description of the problem, the requirements and constraints, and the
IO examples.

At the problem level, every problem has a CSV file. The metadata for
each submission is summarized in Table~\ref{tbl:metadata-problem-level} below, which lists the fields contained in each CSV file as well
as the corresponding descriptions.

\subsection{How to read the CodeNet dataset}
\label{sec:codenet-tour}

The data and metadata are organized in a rigorous directory structure. The top level Project\_CodeNet directory contains several sub-directories: \url{data}, \url{metadata},  \url{problem\_descriptions}, and \url{derived}. The code samples or submissions reside under the data directory. The data directory is organized as
\url{(problem_id)/(language)/(submission)},
so the file path
\url{data/p00023/C++/s006384060.cpp}
denotes a submission to problem p00023 in C++ with id s006384060. Detailed statement of the problems can be found in
\url{problem\_descriptions/(problem_id).html}. 
The meta data for the dataset is contained in the \url{metadata} directory. \url{metadata/problem_list.csv } contains metadata for all the problems in the dataset, which is summarized in \Cref{tbl:metadata-dataset-level}.
\url{metadata/(problem_id).csv} contains the metadata for all the submissions to problem \url{problem_id}, which is described in \Cref{tbl:metadata-problem-level}.
Each submission comes with cpu time, memory usage and status with possible values described in 
\Cref{tbl:all-status-values}.
The \url{derived} directory contains information derived from the dataset, such as  near-duplicate information for submissions to specific languages, token sequences for code samples, and information on identical problems.
\begin{table}[h]
\caption{Metadata at the dataset level}
\label{tbl:metadata-dataset-level}
\centering
\begin{tabular}{|l|l|l|l|}
\hline
\multicolumn{1}{|c|}{\bf{name of column}} & \multicolumn{1}{c|}{\bf{data type}} & \multicolumn{1}{c|}{\bf{unit}} & \multicolumn{1}{c|}{\bf{description}}              \\ \hline
id                                   & string                         & none                      & unique anonymized id of the problem           \\ \hline
name                                 & string                         & none                      & short name of the problem                     \\ \hline
dataset                              & string                         & none                      & original dataset, AIZU or AtCoder             \\ \hline
time\_limit                          & int                            & millisecond               & maximum time allowed for a submission         \\ \hline
memory\_limit                        & int                            & KB                        & maximum memory allowed for a submission       \\ \hline
rating                               & int                            & none                      & rating, i.e., difficulty of the problem       \\ \hline
tags                                 & string                         & none                      & list of tags separated by "|"; not used       \\ \hline
complexity                           & string                         & none                      & degree of difficulty of the problem; not used \\ \hline
\end{tabular}
\end{table}

\begin{table}[h]
\caption{Metadata at the problem level}
\label{tbl:metadata-problem-level}
\centering
\begin{tabular}{|l|l|l|l|}
\hline
\multicolumn{1}{|c|}{\textbf{name of column}} & \multicolumn{1}{c|}{\textbf{data type}} & \multicolumn{1}{c|}{\textbf{unit}} & \multicolumn{1}{c|}{\textbf{description}}   \\ \hline
submission\_id                                & string                                  & none                               & unique anonymized id of the submission      \\ \hline
problem\_id                                   & string                                  & none                               & anonymized id of the problem                \\ \hline
user\_id                                      & string                                  & none                               & anonymized user id of the submission        \\ \hline
\multirow{2}{*}{date}                         & \multirow{2}{*}{int}                    & \multirow{2}{*}{seconds}           & date and time of submission in the Unix     \\ 
                                              &                                         &                                    & timestamp format (seconds since the epoch)  \\ \hline
\multirow{2}{*}{language}                     & \multirow{2}{*}{string}                 & \multirow{2}{*}{none}              & mapped language of the submission           \\ 
                                              &                                         &                                    & (ex: C++14 -\textgreater C++)               \\ \hline
original\_language                            & string                                  & none                               & original language specification             \\ \hline
\multirow{2}{*}{filename\_ext}                & \multirow{2}{*}{string}                 & \multirow{2}{*}{none}              & extension of the filename that indicates    \\ 
                                              &                                         &                                    & the programminglanguage used                \\ \hline
status                                        & string                                  & none                               & acceptance status, or error type            \\ \hline
cpu\_time                                     & int                                     & millisecond                        & execution time                              \\ \hline
memory                                        & int                                     & KB                                 & memory used                                 \\ \hline
code\_size                                    & int                                     & bytes                              & size of the submission source code in bytes \\ \hline
accuracy                                      & string                                  & none                               & number of tests passed; *Only for AIZU      \\ \hline
\end{tabular}
\end{table}

\begin{table}[h]
\caption{All the possible status values}
\label{tbl:all-status-values}
\centering
\begin{tabular}{|l|l|l|}
\hline
\multicolumn{1}{|c|}{\textbf{status}} & \multicolumn{1}{c|}{\textbf{abbreviation}} & \multicolumn{1}{c|}{\textbf{numeric code}} \\ \hline
Compile Error                         & CE                                         & 0                                          \\ \hline
Wrong Answer                          & WA                                         & 1                                          \\ \hline
Time Limit Exceeded                   & TLE                                        & 2                                          \\ \hline
Memory Limit Exceeded                 & MLE                                        & 3                                          \\ \hline
Accepted                              & AC                                         & 4                                          \\ \hline
Judge Not Available                   & JNA                                        & 5                                          \\ \hline
Output Limit Exceeded                 & OLE                                        & 6                                          \\ \hline
Runtime Error                         & RE                                         & 7                                          \\ \hline
WA: Presentation Error                & PE                                         & 8                                          \\ \hline
Waiting for Judging                   & WJ                                         &                                            \\ \hline
Waiting for Re-judging                & WR                                         &                                            \\ \hline
Internal Error                        & IE                                         &                                            \\ \hline
Judge System Error                    &                                            &                                            \\ \hline
\end{tabular}
\end{table}

\begin{figure}[h]
\centering
\includegraphics[height=4.5cm]{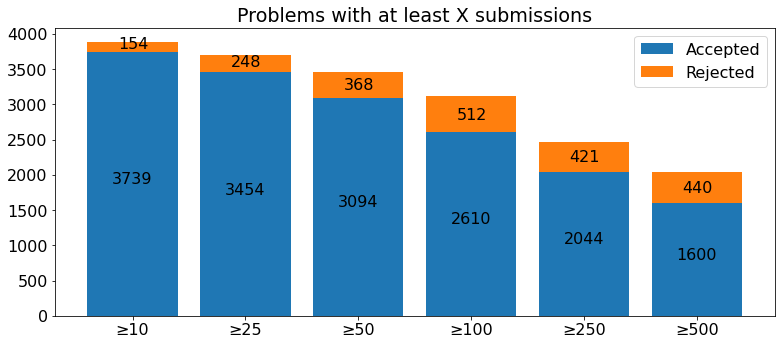} 
  \caption{Number of problems providing at least X submissions. The bars show both the numbers of accepted submissions (blue) and rejected submissions (orange).}
  \label{fig:problems_per_submissions}
\end{figure}

Table~\ref{table:metadata} summarizes the metadata available for each
code submission to a
problem. Figure~\ref{fig:problems_per_submissions} gives the
distributions of problems based on number of submissions received.

\begin{table}[h]
\small
\caption{Submission metadata.}
\label{table:metadata}
\centering
\begin{tabular}{|l|l|l|l|}
\hline
column         & unit/example & description \\
\hline
submission\_id & s[0-9]\{9\} & anonymized id of submission \\
problem\_id    & p[0-9]\{5\} & anonymized id of problem \\
user\_id       & u[0-9]\{9\} & anonymized user id \\
date           & seconds & date and time of submission \\
language       & C++     & consolidated programming language \\
original\_language & C++14 & original language \\
filename\_ext  & .cpp      & filename extension \\
status         & Accepted & acceptance status, or error type \\
cpu\_time      & millisecond & execution time \\
memory         & kilobytes   & memory used \\
code\_size     & bytes       & source file size \\
accuracy       & 4/4         & passed tests (AIZU only) \\
\hline
\end{tabular}
\end{table}

\textbf{Limitations.}
All code samples in CodeNet may not be extensively commented, and these comments may be in multitude of languages. Therefore, AI techniques that rely on learning from preponderance of comments in the code may face challenges. The code samples are solutions to high-school and beginning college level programming problems. This dataset is not suitable for users looking for code with enterprise API's and advanced design patterns.
\section{Related Datasets}
\label{related_dataset}
A wide variety of datasets for source code exist, with many targeting one or a small number of tasks. Such tasks include clone detection,
vulnerability detection~\cite{D2A,Devign}, cloze test~\cite{CTmaxmin},
code completion~\cite{JavaCorpus,Py150}, code repair~\cite{Bugs2Fix},
code-to-code translation, natural language code
search~\cite{CodeSearchNet}, text-to-code generation~\cite{CONCODE},
and code summarization~\cite{CodeSearchNet}. A detailed discussion of
several of these tasks and their respective datasets is available
in CodeXGLUE~\cite{codexglue}, which is a collection of existing datasets. CodeNet, on the other hand, is a new dataset curated from scratch, that aims to support a broad set of 
use cases. Popular datasets of a similar kind are POJ-104~\cite{ncc} (which is incorporated as  part of CodeXGLUE as well)
and GCJ~\cite{gcj} (derived from Google Code Jam). We compare CodeNet to these datasets in the following.

\subsection{POJ-104}
POJ-104 was collected from a pedagogical online judge system. The code
samples are submissions to 104 programming problems. With 500
submissions to each problem, there is a total of 52,000 code samples
in the dataset. This dataset has been used by many authors for
code classification~\cite{ncc} and code similarity~\cite{misim}.

POJ-104 is faced with several limitations.
\begin{enumerate}[leftmargin=*]
\item
The code samples are in C and C++, but the two languages are not
distinguished. Although they are closely related, mixing them leads to
parsing errors and a reduction of useful code samples~\cite{misim}.
\item
Useful metadata such as the results of the judging
system (acceptance, error types etc.) are missing.  Therefore, for certain applications where compilabilty or code
correctness is important, additional pre-processing efforts are needed
and useful code samples are reduced~\cite{misim}.
The dataset does not contain the problem statement, although some
example problems are described in~\cite{poj-problem-samples}, and information on how to execute the code samples is absent.
\item
Some problems are identical (e.g., problems 26 and 62), and some submissions are near duplicates of each other, although the
percentage of such cases is low compared to other datasets.
\end{enumerate}

\subsection{GCJ}
GCJ~\cite{gcj} was collected from the submissions to the Google Code Jam
competitions from 2008 to 2020. Similar to CodeNet, the submissions
cover a wide variety of programming languages, with C++, Java, Python,
and C being the predominant ones. The C++ subset has been extracted
into a POJ-104-like benchmark and used in some
publications. This benchmark dataset, GCJ-297~\cite{gcj-297}, has 297 problems and approximately 280K
submissions. The number of submissions is imbalanced among problems.

GCJ is advantageous over POJ-104 in size and language diversity, but
we believe that an even larger dataset such as CodeNet can better
serve the community. GCJ contains neither metadata nor information on identical problems and near
duplicates.

\section{CodeNet Differentiation}
\label{differentiation}
\begin{table*}[h]
\caption{Related datasets comparison}
\label{tbl:comparison}
\begin{center}
\vskip -10pt
\begin{tabular}{|l|r|r|r|}
\hline
                                        & CodeNet       & GCJ   & POJ \\
\hline
Total number of problems                            &4053           &   332    & 104 \\
Number of programming languages                     &55             &   20    &  2\\
Total number of code samples                        &13,916,828     &   2,430,000    &52,000  \\
C++/C subset data size (code samples)                 &8,008,527      &    280,000   &  52,000\\
Percentage of problems with test data                          & 51\%          &   0\%    &  0\% \\
Task: Memory Consumption Prediction                 & Yes           & No    & No \\
Task: Runtime Performance Comparison                & Yes           & No    & No \\
Task: Error Prediction                              & Yes           & No    & No \\
Task: Near duplicate prediction                     & Yes           & No    & No \\
\hline
\end{tabular}
\end{center}
\end{table*}


A high quality code dataset has certain desired properties. We constructed
CodeNet according to these requirements. In the following, we discuss
how CodeNet differentiates itself from the existing datasets along
these lines. \Cref{tbl:comparison} is a comparison with related datasets.

\textbf{Large scale.}
A useful dataset should contain a large number and variety of data
samples to expose the realistic and complex landscape of data
distributions one meets in practice. CodeNet is the largest dataset in
its class - it has approximately 10 times more code samples than GCJ
and its C++ benchmark is approximately 10 times larger than POJ-104.

\textbf{Rich annotation.}
For the dataset class in question, it is important to include information beyond which problem a code
sample solves to enable a wide range of applications and use cases. It is useful to know
whether a code sample solves the problem correctly, and if not, the
error category (e.g., compilation error, runtime error, and
out-of-memory error). Since the source code is supposed to solve a
programming problem, it is advantageous to know the problem statement
and have a sample input for execution and a sample output for
validation. All such extra information is part of CodeNet but absent in
GCJ and POJ-104.

\textbf{Clean samples.}
For effective machine learning, the data samples are expected to be
independent and identically distributed (iid); otherwise, the
resulting performance metric could be significantly
inflated~\cite{10.1145/3359591.3359735}. The existence of duplicate
and/or near duplicate code samples makes the iid assumption
dubious. Hence, it is crucial to identify the near duplicates. The
presence of identical problems in the dataset poses an even bigger
issue. In CodeNet, we analyzed the code samples for (near) duplication
and used clustering to find identical problems. This information is
made available as part of the dataset release but it is
absent in GCJ and POJ-104.

\section{Construction of CodeNet}

\subsection{Collection of Code Samples}
The CodeNet dataset contains problems, submissions, and metadata,
scraped from the AIZU and AtCoder online judging systems. 
For AIZU, we used the provided REST APIs to download all the metadata. For AtCoder, due to the
absence of a REST API, we scraped the problems, submissions, and
metadata directly from the web pages. 
We considered only public and non-empty submissions that did not contain errors or inconsistencies in the metadata. We manually merged the information from the two sources and adopted a unified format to create a single dataset.

\subsection{Cleansing}
\label{cleansing}
Because data are collected from different sources, we apply a
consistent character encoding (UTF-8) on all raw data
files. Additionally, we remove byte-order marks and use Unix-style
line-feeds as the line ending.

As indicated in section~\ref{differentiation}, we identify near-duplicates.  We follow Allamanis~\cite{10.1145/3359591.3359735} and
use Jaccard similarity~\cite{jaccard} as a metric to score code pairs.
Each code sample is tokenized and stored as a bag-of-tokens
multiset. In our case, we keep all tokens except comments and
preprocessor directives.  We compute the set and multiset Jaccard
indices and respectively use 0.9 and 0.8 as the near-duplicate
thresholds.

\begin{figure}[h]
\centering
\vskip -.2in
\includegraphics[width=\linewidth]{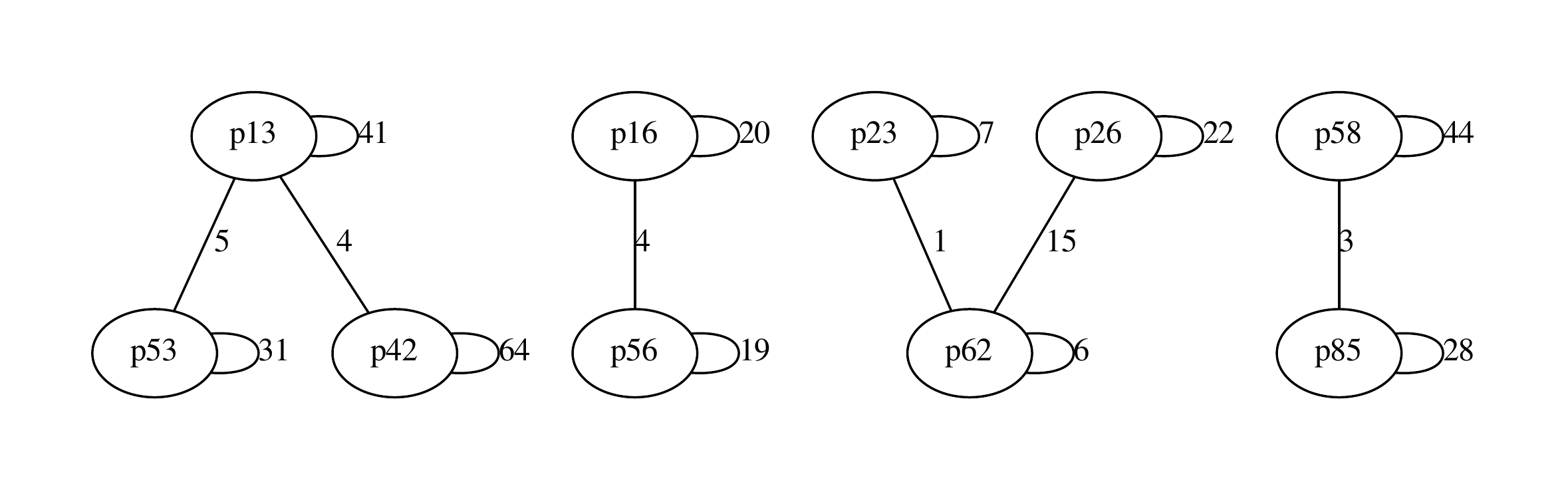}
\vskip -.2in
\caption{An example of a near-duplicate problem graph.}
\label{POJ_104_ident}
\end{figure}

Besides similar code samples, identical problems are also likely because they have been gathered over many decades. We go
through the problem description files (in HTML format) and
apply \code{fdupes} to extract identical problem pairs. Additionally,
using the near-duplicate information calculated for code samples, we
consider a problem pair to be a potential duplicate when the
number of near-duplicate code pairs exceeds a threshold. Clustering of duplicate problems is illustrated by the graphs in
Figure~\ref{POJ_104_ident}, where each node denotes a problem and an
edge between two nodes is labeled by the number of near-duplicate code pairs. Each connected graph is then a cluster of potential duplicate problems and we manually inspect
the problem descriptions to verify the correctness of this duplicate detection.

\subsection{Benchmark Datasets}
CodeNet has a rich set of code samples, and the user can assemble a
customized benchmark according to his/her need. Following POJ-104, we
extracted benchmark datasets from CodeNet in C++, Python, and Java. 
The benchmark characteristics are shown in \Cref{tbl:spt-software-artifacts}.
For the C++ benchmarks, the number of problems and their solutions are chosen to make the benchmark challenging. The benchmarks are filtered in the following ways.
Each code sample is ``unique'' in the sense that it is not a
near-duplicate of another code sample.
The same is true of each problem.
Samples with a large fraction of dead code are excluded.
Each code sample has successfully passed through the tokenizer, the SPT
generator, and the graph generator, all described in the next
section. This step is to ensure that proper processing can be done to
convert a code sample to a machine learning model input.

\section{Code Representation and Tools}
\label{repr}

Machine learning with source code requires proper abstractions of the code. The abstractions are instantiated as representations in specific formats. As a usability feature, we provide several pre-processing tools to transform source codes into representations that can readily be used as inputs into machine learning models. They are described as follows.

\textbf{Tokenizer.}
We offer fast C implementations
of tokenizers for C, C++, Java, Python, and JavaScript. Additionally,
the parse-tree generator described next can also produce token streams
for C, C++, Java, and Python and can easily be extended to more
languages.

\label{sec-spt}


\textbf{Simplified Parse Tree (SPT)}
Simplified parse trees are derived from parse trees generated using ANTLR4 \cite{10.5555/2501720}. We traverse the ANTLR4 parse tree and remove internal nodes that only have one child. 
By doing so, we maintain the essential structure of
the parse tree while pruning out unnecessary parser production rules. Finally, we adopt Aroma's~\cite{aroma}
naming convention: leaf nodes are named by their
literal strings and internal nodes are named by a concatenation of
their children's names (only reserved words are kept while others are
replaced by a hash mark $\#$).
We produce features for each node: (1) node type (token or parsing
rule); (2) token type (e.g., an identifier), when applicable; (3)
parsing rule type (e.g., an expression), when applicable; and (4)
whether it is a reserved word. We adopt an
extensible JSON graph schema so that edges can be augmented with types
when needed.
Currently, we support generating SPTs for four languages: C, C++,
Java, and Python. \Cref{tbl:spt-software-artifacts} summarizes the SPT statistics for the four benchmarks. 

\begin{table}[h]
\small
\caption{Benchmark statistics.}
\label{tbl:spt-software-artifacts}
\centering
\begin{tabular}{|l|r|r|r|r|}
\hline
                & C++1000 & C++1400 & Python800     & Java250       \\
\hline
\#problems       & 1,000       & 1,400       & 800        & 250        \\
\#samples       & 500,000     & 420,000     & 240,000    & 75,000     \\
\#SPT-nodes         & 188,449,294  & 198,258,050  & 55,744,550 & 25,449,640 \\
\#SPT-edges         & 187,949,294  & 197,838,050  & 55,504,550 & 25,374,640 \\
\hline
\end{tabular}
\end{table}

\textbf{Code graphs.} We augment the tool chain with a code
graph generator using WALA \cite{wala}, a general framework for program
analysis.
The backbone of a code graph is a system dependence graph, which is an inter-procedural graph of program instructions (e.g. call, read) expressing control flow and data flow information as edges.
We also generate inter-procedural control flow graphs, which are control flow graphs of all the methods in the program, stitched
together to connect call sites with target methods.
Our code graph tool currently supports only Java and Python, but we plan to support more languages such as Javascript.

\section{CodeNet Challenge}
\label{contest}
The launch of CodeNet was well received by the AI community and the media, with coverage from Forbes\cite{forbes}, VentureBeat\cite{venturebeat}, ZDNet\cite{zdnet} and others. Within a short span of 3 months, our github received 1000 stars and has been forked over 119 times. Our vision is to use CodeNet as an umbrella to curate AI for code datasets for widespread adoption and to drive innovation in AI for code. To leverage the momentum of CodeNet, we will be launching CodeNet challenges to create excitement in the AI community. The first contest ~\cite{tier1contest} is mainly pedagogical and targets aspiring data scientists. In addition, we are partnering with the Global Women in Data Science organization (with presence in over 50 countries) founded by Stanford University \cite{wids} to emphasize diversity and inclusion (teams must have at least fifty percent women). We will organize workshops to introduce the topic, code similarity, and provide educational materials. This contest will be kicked off in late September and the winner will be announced in early December, around the NeurIPS2021 time frame. The conclusion of the first contest will be followed by a contest that will target experienced AI practitioners. Potential contest topics will revolve around practical and compelling use cases such as code language translation, code repair, code performance improvement, and code memory reduction.

\section{Experiments with the CodeNet Dataset}
\label{baseline_expts}
In this section, we report the results of a code classification task, a similarity task, a generalization task, and a token inference task, using the four benchmark datasets (see \Cref{tbl:spt-software-artifacts}) extracted from
CodeNet. For this paper, these experiments are not meant to achieve the best-of-breed results using the state of the art. Our intention is to provide a set of 
baseline results as a reference. The experiments are typically performed on a Xeon machine using P100 or V100 GPUs. Code and scripts for these experiments are in the model-experiments folder of the CodeNet repository~\cite{projectcodenet}.

\subsection{Code Classification}
In the classification task, each problem corresponds to a class: a code sample
belongs to a class if it is a submission to the corresponding
problem. For each
experiment, 20\% of the code samples are used for testing, while the
rest are split in 4:1 for training and validation, respectively.
We experiment with a diverse set of machine learning methods: bag of tokens, sequence of tokens, BERT model, and graph neural networks (GNNs). 

\begin{enumerate}[leftmargin=*]
\item \textbf{MLP with bag of tokens.}
A code sample is represented by a vector of relative frequencies of
token occurrences. Only operator and keyword tokens are used. The
model is a 3-layer multilayer perceptron (MLP).

\item \textbf{CNN with token sequence.}
We use the same set of tokens as above but retain their order to form
a sequence. All sequences have the same length under zero padding. The
classification model is a convolutional neural network (CNN) with an initial token embedding layer.

\item \textbf{C-BERT with token sequence.}
Treating a code sample as a piece of natural language text, we build a
C-BERT model~\cite{cbert} through pretraining on 10K top starred
Github projects written in C. We use the Clang C tokenizer and
Sentencepiece to tokenize each code sample. The pretrained model is
fine-tuned on each benchmark.

\item \textbf{GNN with SPT.}
Based on the parse tree representation, we use graph convolutional
networks (GCN)~\cite{Kipf2017} and graph isomorphism networks (GIN)~\cite{Xu2019} as well as  their variants
as the prediction model. The variant adds a virtual
node to the graph to enhance graph message passing~\cite{Thost2021}.

\item \textbf{GNN with Code  Graph.}
We also apply GCN on the code graph representation of the code.
\end{enumerate}

\begin{table}[h]
  \small
  \caption{Classification accuracy (in \%).}
  \label{table:code.classification}
  \centering
  \begin{tabular}{|l|r|r|r|r|}
    \hline
    & Java250 & Python800 & C++1000 & C++1400 \\
    \hline
    MLP w/ bag of tokens  & 71.00$\pm$0.29 & 67.80$\pm0.15$ & 68.26$\pm$0.21 & 64.50$\pm$0.13 \\
    CNN w/ token sequence & 89.52$\pm$0.59 & 87.46$\pm$0.25 & 93.96$\pm$0.18 & 93.71$\pm$0.18 \\
    C-BERT                & 97.40$\pm$0.19 & 97.09$\pm$0.18 & 93.79$\pm$0.01 & 91.83$\pm$0.06 \\
    GNN (GCN)             & 92.70$\pm$0.25 & 93.82$\pm$0.16 & 95.76$\pm$0.12 & 95.26$\pm$0.13 \\
    GNN (GCN-V)           & 93.02$\pm$0.81 & 94.30$\pm$0.15 & 96.09$\pm$0.17 & 95.73$\pm$0.07 \\
    GNN (GIN)             & 93.26$\pm$0.23 & 94.17$\pm$0.19 & 96.34$\pm$0.15 & 95.95$\pm$0.13 \\
    GNN (GIN-V)           & 92.77$\pm$0.66 & 94.54$\pm$0.12 & 96.64$\pm$0.10 & 96.36$\pm$0.10 \\
    Code Graph+GCN & 94.10$\pm$.001 & 87.80$\pm$.007 & N/A & N/A \\
    \hline
  \end{tabular}
\end{table}

Table~\ref{table:code.classification} summarizes the classification
accuracy for all models on all benchmarks. Despite the simplicity of
bag of tokens, it achieves well over 60\%
accuracy. Maintaining token  ordering, CNN with token sequence offers significant improvement, reaching approximately 90\%
across all benchmarks.

More complex neural models sometimes further improve the prediction
performance, as witnessed by C-BERT, which reaches approximately 97\%
for both Java and Python. It is interesting to note that even though C-BERT is
pre-trained with C programs, its performance on the two C++
benchmarks is less impressive. We speculate that such a lower
performance is related to programming practices. 
For C++, it is common to have identical program construction, such as declaration of constants (e.g., pi and epsilon) and data structures,
appear across C++ submissions to different problems, but such
a practice is rare in Java and Python.

Overall, the GNN models exhibit competitive performance. They are consistently the top performers, if not the best.
The code graph representation slightly improves over the SPT representation on Java, but performs less well on Python.

Further details of each model, along with the experiment environment, are given below.
\subsubsection{Details of Experiments on Code Classification}
\label{appendix:code_class_details}

\textbf{MLP with Bag of Tokens}\\
\label{bag_tok_class}
One of the simplest representations of a code sample is a bag of
tokens. Here, the code sample is represented by a vector of relative
frequencies of token occurrences in the source code. The vector is
computed by the following steps:
\begin{enumerate}[leftmargin=*]
\item
Convert a given source code into a sequence of tokens using a
tokenizer (i.e., lexical analyzer).
\item
From this sequence, remove the tokens considered not useful for code
classification.
\item
Count the number of each token type in the reduced sequence and form a
vector of counts.
\item
Normalize the vector with respect to L2 norm.
\end{enumerate}

We do not use all tokens available in the grammar of the programming
language. Only some operators and keywords are used. All identifiers,
comments and literals are ignored. We also ignore some operators and
many keywords that in our opinion provide no significant information
on the algorithm the source code implements.

The vector representing a bag of tokens has the same length for every
code sample, which makes it convenient for processing with a neural
network. The vector is usually short, which makes training of a neural
network fast. However, in a bag-of-tokens representation, information
about the number of occurrences and position of each token is
lost. Hence, the accuracy of a classifier using a bag-of-tokens
representation is rather limited.

Table~\ref{table:bag_class} provides results of code classification of
all four benchmarks. The columns give the benchmark name, the test
accuracy, the number of training epochs, the run time of each epoch,
and the number of token types considered. All networks are implemented
using Keras API of TensorFlow machine learning tool. 
Training is performed on a single V100 GPU, using Adam optimizer with 
learning rate 1e-3, and batches of 32 samples.
In each experiment, 20\% of the 
samples are used for testing, while the
rest are split in 4:1 for training and validation, respectively.
 
\begin{table}[h]
\small
\caption{Code classification by MLP with bag of tokens.}
\label{table:bag_class}
\centering
\begin{tabular}{|l|r|r|r|r|}
\hline
Benchmark & Accuracy & Number & Run time  & Number \\
dataset   & \%\%        & epochs   & sec/epoch & tokens \\
\hline
 Java250  & 71.00$\pm$0.29 & 30 & 2  &   81   \\
 Python800& 67.80$\pm$0.15 & 22 & 7  &   71   \\
 C++1000  & 68.26$\pm$0.21 & 20 & 14 &   56   \\
 C++1400  & 64.50$\pm$0.13 & 17 & 12 &   56   \\
\hline
\end{tabular}
\end{table}

\begin{figure}[h]
\centering
\includegraphics[scale=0.75]{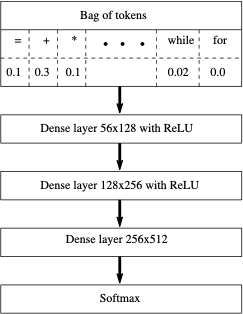}
\caption{MLP architecture for code classification.}
\label{bag_tok_class_fig}
\end{figure}

Figure~\ref{bag_tok_class_fig} shows the neural network used for
solving the classification problem for the C++1400 benchmark. 
The neural networks used for classification of other benchmarks are
similar to this one. 
As we see in Table~\ref{table:bag_class} their performance is quite similar.

From Table~\ref{table:bag_class} we see that training is rather fast,
the reason being that the network is simple. In spite of simplicity,
this neural network performs very well. 
The 64.50$\pm$0.13\% test accuracy for C++1400 benchmark dataset is
significantly better than the potential 0.071\% accuracy of random
guess. It indicates that the relative frequencies of source code
tokens provide sufficient information for classifying code.

\textbf{CNN with Token Sequence}\\
\label{seq_tok_class}
The sequence-of-tokens representation retains more information of a
code sample than the bag-of-tokens representation. For our experiments
on code classification, we use the same set of tokens that is used in
the above bag-of-tokens approach. Similarly, we omit all comments and
identifiers.

\begin{table}[h]
\small
\caption{Code classification by CNN with token sequence.}
\label{table:seq_class}
\centering
\begin{tabular}{|l|r|r|r|r|}
\hline
Benchmark & Accuracy & Number & Run time  & Number \\
dataset   & \%\%     & epochs & sec/epoch & tokens \\
\hline
 Java250  & 89.52$\pm$0.59 & 810 & 10  &  81   \\
 Python800& 87.46$\pm$0.25 & 504 & 26  &  71   \\
 C++1000  & 93.96$\pm$0.18 & 235 & 59  &  56   \\
 C++1400  & 93.71$\pm$0.18 & 334 & 60  &  56   \\
\hline
\end{tabular}
\end{table}

Table~\ref{table:seq_class} shows results of code classification on
all four benchmarks by using the sequence-of-tokens
representation. The columns give the benchmark name, the test
accuracy, the number of training epochs, the run time of each epoch,
and the number of token types considered. All networks are implemented
using Keras API of TensorFlow machine learning tool. 
The training is performed on four V100 GPUs, using Adam optimizer 
in data parallel mode with learning rate 1e-3, and batches of 512 samples.
In each experiment, 20\% of the 
samples are used for testing, while the
rest are split in 4:1 for training and validation, respectively.

\begin{figure}[h]
\centering
\includegraphics[scale=0.75]{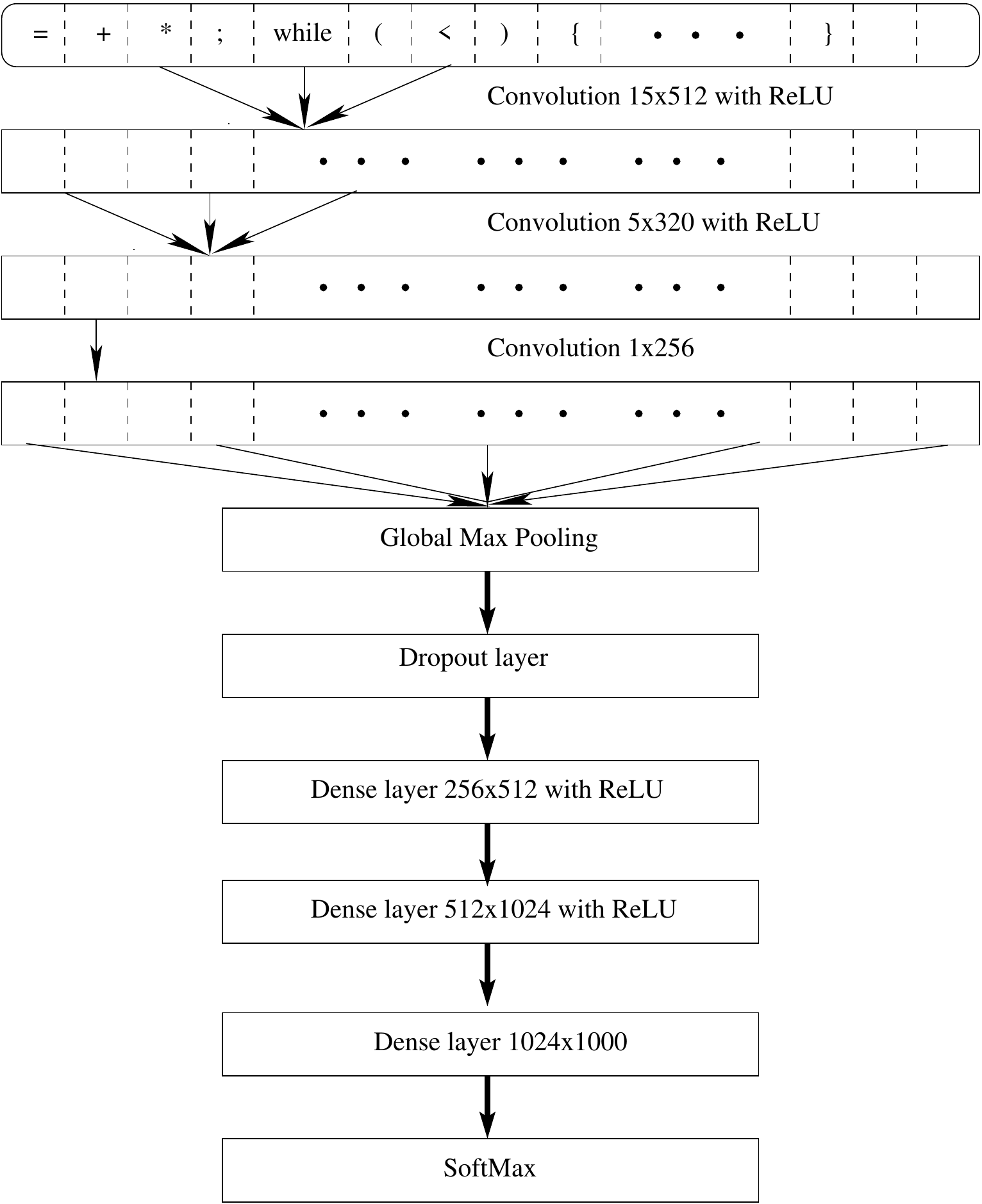}
\caption{CNN architecture for code classification.}
\label{seq_tok_class_fig}
\end{figure}

We have experimented with several types of neural
networks. Figure~\ref{seq_tok_class_fig} shows the neural network we
choose for the C++1400 benchmark. It is a multi-layer convolutional
neural network. It uses categorical encoding of source code
tokens. For batching, the sequences of tokens are padded with zeros.

Using this network we get a test accuracy 93.71$\pm$0.18\% for C++1400 benchmark dataset, which is
significantly better than the accuracy shown by the bag-of-tokens
approach.
The neural networks used for classification of other benchmarks are
similar to the one shown in Figure~\ref{seq_tok_class_fig}.  As we see
in Table~\ref{table:seq_class}, their performance is similar.

\textbf{C-BERT with Token Sequence}\\
The sequence-of-tokens representation can be used with other neural
networks of increasing capacity. We build a C-BERT model (a
transformer model introduced in~\cite{cbert}) by pre-training on
10,000 top starred GitHub open source projects written in C, where we
use Clang C tokenizer and Sentencepiece to tokenize the pre-training
data. The C-BERT model is then fine tuned on each classification
benchmark. Additionally, we experiment with the POJ-104 dataset, which
contains code examples in C and C++.

C-BERT achieves appealing results on binary classification and
vulnerability detection with C source
code~\cite{D2A,DBLP:journals/corr/abs-2102-04664}.  However, it has
not been used on multiclass classification tasks or with other
languages such as C++, Java, and Python.  Because we use sub-word
tokenization and different programming languages share common tokens,
we could apply the C-BERT model directly on the benchmarks.

After pretraining, we fine tune the model for five epochs on each
benchmark, with a batch size 32 and learning rate 2e-5. The
fine-tuning was done on two V100 GPUs and it took 30 minutes to four
hours, depending on the size of the dataset.  The sub-word vocabulary
size is 5,000. Contexts larger than 512 tokens were truncated.

\Cref{tbl:cbert} summarizes the accuracies C-BERT achives on the four
CodeNet benchmarks as well as the POJ-104 dataset.  C-BERT achieves
high accuracy and performs the best on Java and Python.

\begin{table}[h]
\small
\caption{C-BERT results (accuracy, in \%) for code classification.}
\label{tbl:cbert}
\centering
\begin{tabular}{|l|c|c|c|c|c|}
\hline
            & POJ-104 & C++1000 & C++1400 & Java250    & Python800 \\
\hline
C-BERT         & 98.41$\pm$0.01        & 93.79$\pm$0.01        & 91.83$\pm$0.06        & 97.40$\pm$0.19 & 97.09$\pm$0.18 \\
\hline
\end{tabular}
\end{table}

The relatively low performance on C++ benchmarks is possibly related
to the idiosyncrasies of the dataset and certain programming
practices.  Manual inspection suggests that lack of detailed variable
names in C++ hurts the performance of the model, in problems appearing
similar and having similar solutions.  Removing one of the similar
problems improves the model performance on the other problem.
Moreover, one programming practice which could potentially confuse the
models is that certain C++ users copied common constants (e.g., pi and
epsilon) and data structures (e.g., enums) to all solutions they
submitted.  In many cases, these duplicate contents were not even
used. We did not observe such practices in Python and Java.

\textbf{GNN with SPT}\\
We experiment with four types of GNNs with SPT-based graph representations of the source code: the Graph Convolutional Network (GCN)~\cite{Kipf2017}, the Graph Isomorphism Network (GIN)~\cite{Xu2019}, and a virtual-node-included variant for each (denoted by -V). The variant adds a virtual node to the graph to enhance graph message
passing~\cite{Thost2021}.  We use the Adam optimizer
with learning rate 1e-3 for training. All GNN models have five layers. We have experimented with more than 5 layers (i.e., 8 and 10), however deeper GNNs do not improve performance, as deeper GNNs might suffer from the over-smoothing problem (i.e., node features become less distinguishable after many rounds of message passing) \cite{li2018deeper}. 

We conduct 6/2/2 random split for each of the 4 benchmarks: i.e., 60\% training data, 20\% testing data, and 20\% validation data.  We run five folds for each benchmark with early stop ''patience'' set 20 (i.e., stop only when validation loss has not decreased in the past 20 epochs). Our model training typically converges within 200 epochs in a 1-fold run. We modified OGB \cite{hu2020ogb} code-base with PyTorch Geometric \cite{pyg} back-end over PyTorch 1.6.0 \cite{pytorch} to run our experiments. The experiments are conducted on one NVIDIA
V100 GPU. For large benchmarks such as C++1000 and C++1400, it takes about 1 week to finish a 5-fold run. We summarize model accuracy, training time over 5-folds, and training epochs over 5-folds in \Cref{tbl:spt_results}. As we can see, adding a virtual node
improves GNN performance (both GCN and GIN). Overall,  GIN and its variants work better than GCN and its variants, likely due to the fact that GIN theoretically generalizes the Weisfeiler-Lehman Isomorphism Test and achieves maximum expressive power among GNNs \cite{xu2019powerful}.
   
For the detailed model, hyper-parameter setup, data splits and etc, please refer to \url{https://github.com/IBM/Project_CodeNet/tree/main/model-experiments/gnn-based-experiments}.

\begin{table}[h]
\small
\caption{GNN (SPT) results for code classification. Each task trains over 5-folds with early stopping patience parameter set as 20. We record test accuracy (with standard deviation), total training time over 5 folds, and total training epochs over 5 folds.  }
\label{tbl:spt_results}
\centering
\begin{tabular}{|l|c|c|c|c|}
\hline
            & Java250 & Python800 & C++1000    & C++1400  \\
\hline
GCN            & 92.70$\pm$0.25   & 93.82$\pm$0.16  & 95.76$\pm$0.12  & 95.26$\pm$0.13  \\
                      & 10.55 hrs                     &    14.50 hrs                      &   47.96 hrs                       & 67.34 hrs \\
                      &  411 epochs                     &  219 epochs                        & 228 epochs                         &  310 epochs \\
                      \hline
GCN-V          & 93.02$\pm$0.81  & 94.30 $\pm$0.15  & 96.09$\pm$0.17  & 95.73$\pm$0.07   \\
                      &  12.50 hrs                       &    23.02 hrs                      &   61.55 hrs                       & 71.85 hrs \\
                      & 419 epochs                         &  325 epochs                        &  287 epochs                        & 358 epochs \\
\hline
GIN           & 93.26$\pm$0.23  & 94.17$\pm$0.19  & 96.34$\pm$0.15  & 95.95$\pm$0.13  \\
                      &   19.80 hrs                       &  41.67 hrs                        &    116.67 hrs                      & 133.50 hrs\\
                      & 513 epochs                         & 496 epochs                         & 441 epochs                         & 502 epochs \\
\hline 
GIN-V           & 92.77$\pm$0.66  & 94.54$\pm$0.12  & 96.64$\pm$0.10  & 96.36$\pm$0.10  \\
                      &  26.25 hrs                        &    51.67 hrs                      &  142.25 hrs                        & 208.47 hrs\\
                      &   656 epochs                      &   570 epochs                       &  496 epochs                        & 678 epochs \\
\hline
\end{tabular}
\end{table}

\subsection{Code Similarity}
In the similarity task, two pieces of code samples are considered similar
if they solve the same problem (type-4 similarity
in~\cite{similarityType}). Note that textual similarity does not
guarantee similarity in functionality. For example,
programs that differ by only one token might behave very differently;
hence, they are not considered similar. For the token-based experiments, we treat the problem as binary classification. 
We use the same training, validation and testing split as in classification.
Code pairs are randomly sampled within each subset. The
number of similar pairs is the same as dissimilar ones.
For the SPT representation, we experiment with several popular techniques, including AROMA~\cite{aroma}, MISIM~\cite{misim}, and GMN~\cite{gmn}.
The following contains more details about the models and methods.

\begin{enumerate}[leftmargin=*]
\item \textbf{MLP with bag of tokens.}
This model is the same as the one for code classification, except that
the input is a concatenation of the two bag-of-tokens vectors from
each program.

\item \textbf{Siamese network with token sequence.}
The token sequence is the same as the one for code classification. The
model is a Siamese network with two CNNs with shared weights.


\item \textbf{SPT with handcrafted feature extraction:} The method AROMA~\cite{aroma} uses normalized SPT node names and handcrafted rules to extract feature vectors for each SPT. Then, similarity is computed as a dot product of the extracted feature vectors. 

\item \textbf{GNN with SPT:} With the same SPT, on the other hand, MISIM~\cite{misim} uses a graph neural network to extract
high-level features, and uses the cosine similarity of the extracted
features to compute similarity. Additionally, we apply graph matching network (GMN)~\cite{gmn}, which uses a cross-graph attention mechanism to learn pair-wise structural similarity of graphs, on the SPT pairs to predict similarity. The implementation is adapted from~\cite{gmnimpl}.

\end{enumerate}

\begin{table}[h]
  \small
  \caption{Similarity accuracy (in \%).}
  \label{table:code.similarity.1}
  \centering
  \begin{tabular}{|l|r|r|r|r|}
    \hline
    & Java250 & Python800 & C++1000 & C++1400 \\
    \hline
    MLP w/ bag of tokens      & 81.80$\pm$0.06 & 86.61$\pm$0.08 & 85.82$\pm$0.05 & 86.54$\pm$0.07 \\
    Siamese w/ token sequence & 89.70$\pm$0.18 & 94.67$\pm$0.12 & 96.19$\pm$0.08 & 96.56$\pm$0.07 \\
    \hline
  \end{tabular}
\end{table}

Table~\ref{table:code.similarity.1} summarizes the classification
accuracy for the first two models. The performance of bag of tokens is
modest, considering that the problem is a binary classification with
perfectly balanced classes. On the other hand, the Siamese model
significantly outperforms bag of tokens, as expected.

\begin{table}[h]
  \small
  \caption{Similarity MAP@R score.}
  \label{table:code.similarity.2}
  \centering
  \begin{tabular}{|l|c|c|c|c|}
    \hline
    & Java250 & Python800 & C++1000 & C++1400 \\
    \hline
    Rule-based w/ SPT (AROMA) & 0.19 & 0.19 & 0.17 & 0.15 \\
    GNN w/ SPT (MISIM) & 0.64$\pm$0.007 & 0.65$\pm$0.003 & 0.78$\pm$0.005 & 0.77$\pm$0.002 \\
    \hline
  \end{tabular}
\end{table}

Table~\ref{table:code.similarity.2} summarizes the MAP@R~\cite{mapr} score for two SPT-based approaches with solutions for 50\% problems used for training, 25\% for validation, and 25\% for test. MISIM GNN model is trained for 1000 epochs. AROMA results in a relatively low score because the feature extraction is rule-based and no model is learned, whereas MISIM uses a neural network to extract features through supervised training. 

\begin{table}[h]
  \small
  \caption{Similarity MAP@R score on Java250.}
  \label{table:code.similarity.3}
  \centering
  \begin{tabular}{|l|c|c|c|c|}
    \hline
    & (p4, s5) & (p3, s300) & (p10, s300) \\
    \hline
    GNN w/ SPT (MISIM, structure only)   & 0.472$\pm$0.023 & 0.194$\pm$0.010 & 0.096$\pm$0.009 \\
    GNN w/ SPT (GMN, structure only)   & 0.679$\pm$0.056 & 0.432$\pm$0.035 & 0.256$\pm$0.015  \\
    GNN w/ SPT (GMN + MISIM node attributes) & 0.985$\pm$0.015 & 0.794$\pm$0.036 & 0.780$\pm$0.026  \\
    \hline
  \end{tabular}
\end{table}
Exploring further into the Java250 benchmark, Table~\ref{table:code.similarity.3} summarizes the MAP@R score with a variety of test sets: (p4, s5), (p3, s300), and (p10, s300), indicating 4, 3, and 10 problems with 5, 300 and 300 solutions each respectively. Across all test sets, GMN outperforms MISIM if both are trained with only the SPT structure; when combined with MISIM node attributes, GMN further improves the score significantly.

\begin{figure}[hbt]
\centering
\includegraphics[scale=0.35]{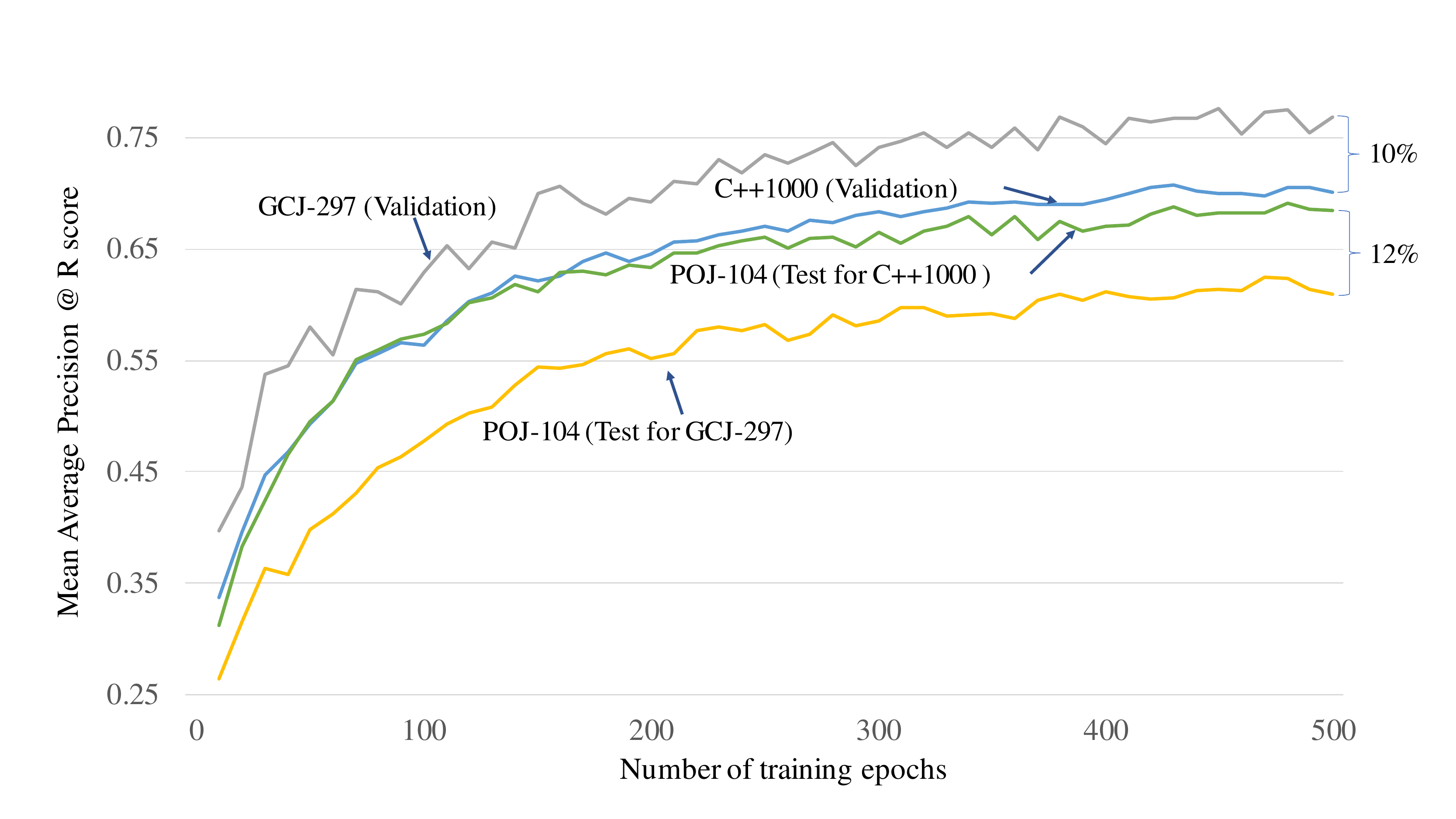}
\caption{Test score on POJ-104 is $12$\% higher when a model is trained on C++1000 as compared to a model trained on GCJ-297, even though the validation score for GCJ-297 model is $10$\% higher than the validation score for C++1000 model.}
\label{codenet_vs_gcj}
\end{figure}

Further details of each model, along with the experiment environment, are given below.

\subsubsection{Details of Experiments on Code Similarity}
\label{appendix:code_sim_details}

\textbf{MLP with Bag of Tokens}\\
\label{bag_tok_sim}
For experiments on code similarity analysis, we use the same bag of
tokens as for code classification. The input to the neural network is
constructed by concatenating two bags of tokens, one for each source
code file.

Table~\ref{table:bag_sim} provides results of code similarity analysis
on all four benchmarks. The columns give the benchmark name, the test
accuracy, the number of training epochs, the number of samples in each
epoch, the run time of each epoch, the number of token types
considered, and the number of test samples. All networks are
implemented using Keras API of TensorFlow machine learning tool. 
The training is performed on a single V100 GPU, 
using Adam optimizer with learning rate 1e-3, and batches of 256 samples.

\begin{table}[h]
\small
\caption{Similarity analysis by MLP with bag of tokens.}
\label{table:bag_sim}
\centering
\begin{tabular}{|l|r|r|r|r|r|r|}
\hline
Benchmark  & Accuracy & Number & Size of & Run time  & Number & N test \\
dataset    & \%\%     & epochs & epoch  & sec/epoch & tokens & samples \\
\hline
 Java250  & 81.80$\pm$0.06 & 20 & 4,096,000 & 21 & 81 & 512,000 \\
 Python800& 86.61$\pm$0.08 & 94 & 4,096,000 & 24 & 71 & 512,000 \\
 C++1000  & 85.82$\pm$0.05 & 64 & 4,096,000 & 21 & 56 & 512,000 \\
 C++1400  & 86.54$\pm$0.07 & 64 & 4,096,000 & 22 & 56 & 512,000 \\
\hline
\end{tabular}
\end{table}

\begin{figure}[h]
\centering
\includegraphics[scale=0.75]{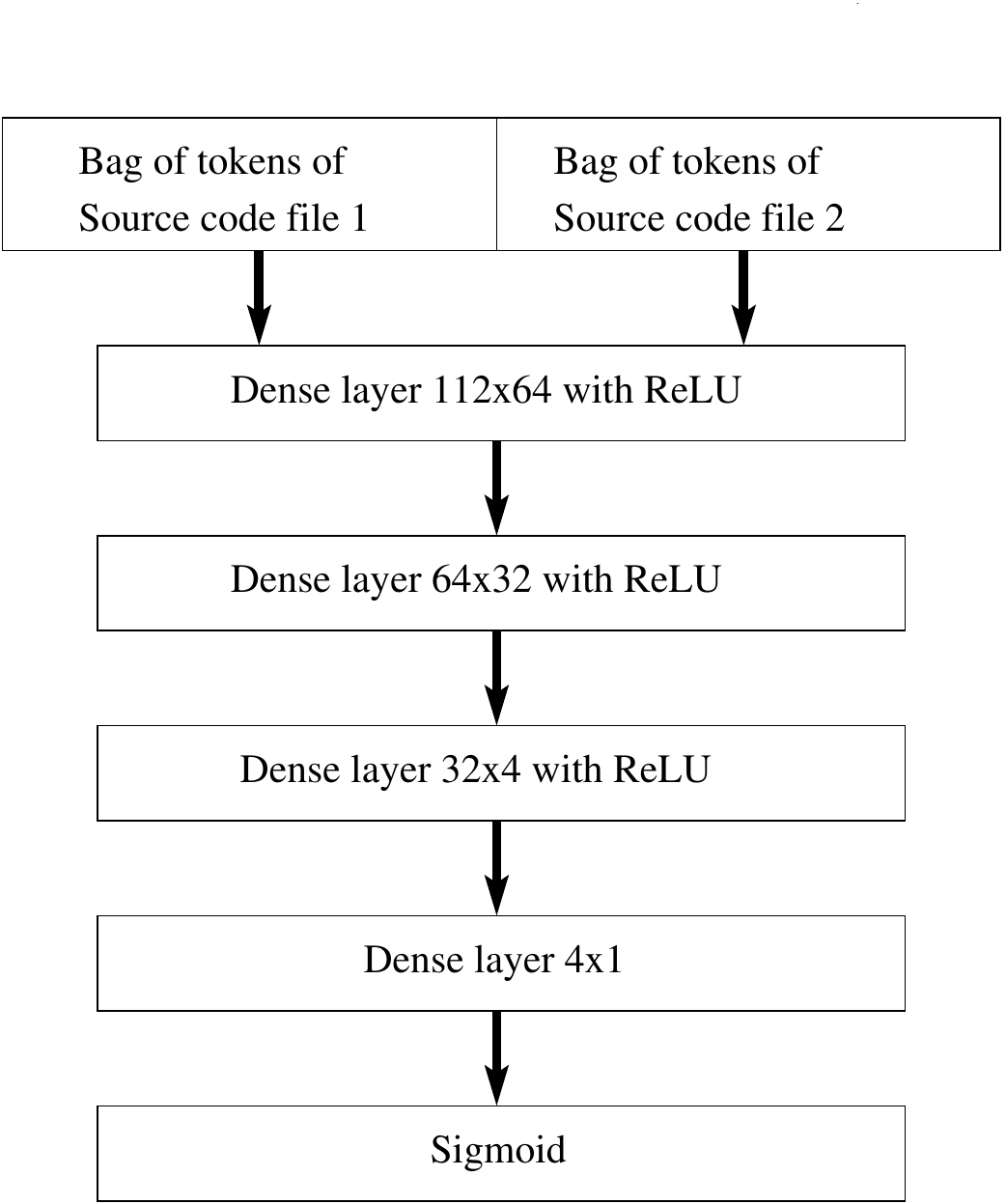}
\caption{MLP architecture for similarity analysis.}
\label{bag_tok_sim_fig}
\end{figure}

Figure~\ref{bag_tok_sim_fig} shows the neural network used for code
similarity analysis on the C++1400 benchmark. 
The neural networks used for code similarity analysis on other
benchmarks are similar to this one.  As we see in Table~\ref{table:bag_sim}, their accuracy is similar.

As we see in Table~\ref{table:bag_sim}, the model accuracy is rather modest
(<87\%) for all benchmark datasets, which is not very high for a binary classification problem of
a fully balanced dataset. Obviously, the bag of tokens is too
primitive and misses many important details necessary for identifying
similarity.

\textbf{Siamese Network with Token Sequence}\\
\label{seq_tok_sim}
For experiments on code similarity, we use the same sequence of tokens as
for code classification. The neural network has two inputs, one for
each source code file. After experimenting with various neural network
architectures, we select the siamese network for its good performance.

Table~\ref{table:seq_sim} provides results of code similarity analysis
on all four benchmarks. The columns give the benchmark name, the test
accuracy, the number of training epochs, the number of samples in each
epoch, the run time of each epoch, the number of token types
considered, and the number of test samples.  All networks are
implemented using Keras API  of TensorFlow machine learning tool. 
The training is performed on four V100
GPUs, using Adam optimizer in data parallel mode with learning rate 1e-3,
 and batches of 512 samples.
 
\begin{table}[h]
\small
\caption{Similarity analysis by Siamese network with token sequence.}
\label{table:seq_sim}
\centering
\begin{tabular}{|l|r|r|r|r|r|r|}
\hline
Benchmark & Accuracy & Number & Size of & Run time  & Number & N test \\
dataset   & \%\%     & epochs & epoch  & sec/epoch & tokens & samples \\
\hline
 Java250  & 89.70$\pm$0.18 & 29  & 51,200 & 114 & 75 & 512,000\\
 Python800& 94.67$\pm$0.12 & 110 & 64,000 & 89 & 71 & 512,000 \\
 C++1000  & 96.19$\pm$0.08 & 123 & 64,000 & 89 & 56 & 512,000 \\
 C++1400  & 96.56$\pm$0.07 & 144 & 64,000 & 96 & 56 & 512,000 \\
\hline
\end{tabular}
\end{table}

\begin{figure}[h]
\centering
\includegraphics[scale=0.75]{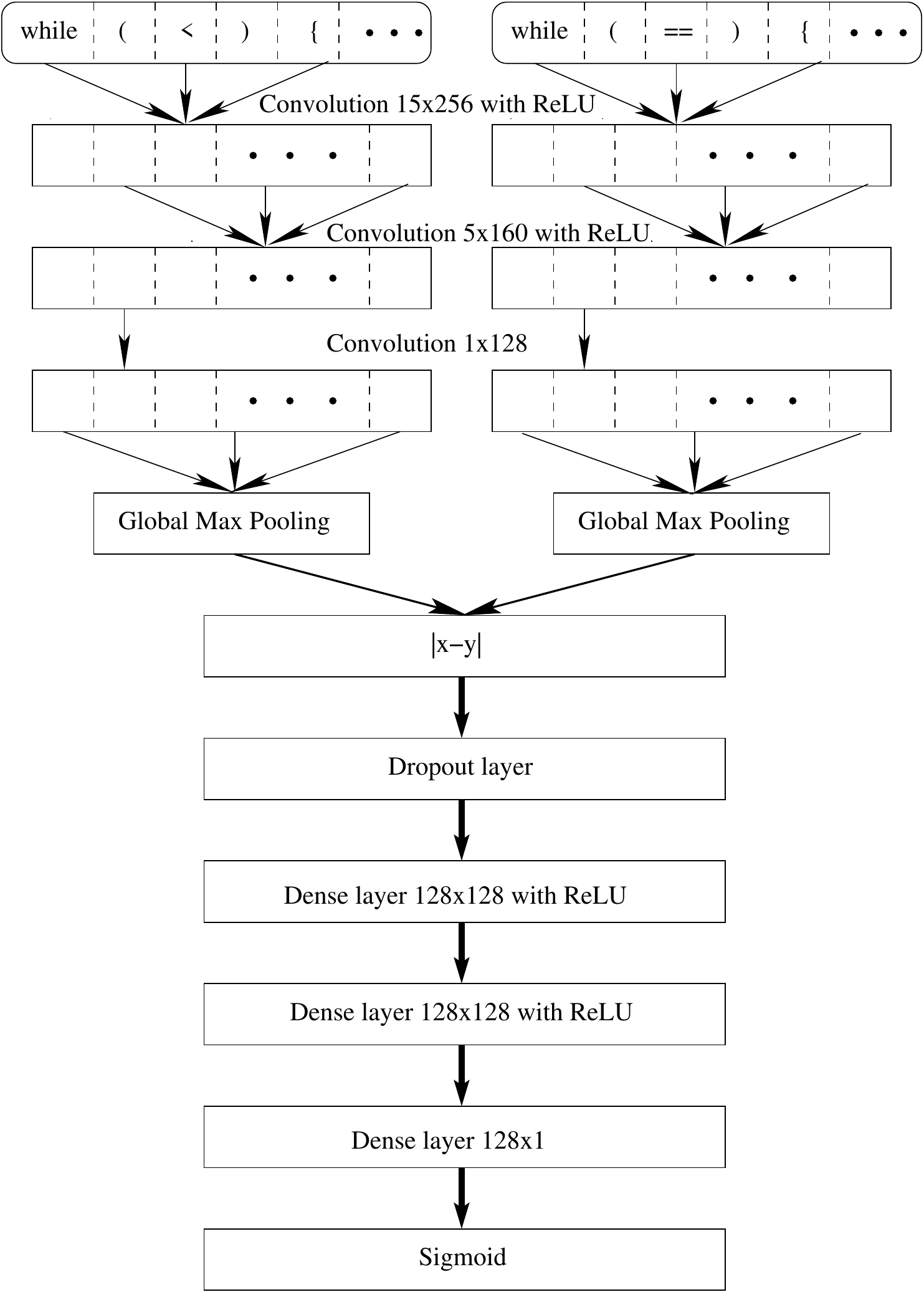}
\caption{Siamese architecture for similarity analysis.}
\label{seq_tok_sim_fig}
\end{figure}

The neural network for the C++1400 benchmark is depicted in
Figure~\ref{seq_tok_sim_fig}. The siamese parts of the network have
the same structure and share all their weights. If the inputs are
identical, so are the outputs. Therefore, by construction, the network
guarantees detecting similarity of identical source code samples.
The outputs of the siamese parts are compared by computing the
absolute difference.

The network shows 96.56$\pm$0.07\% test accuracy for C++1400 benchmark dataset. We consider this a good
result, especially considering that the token sequence ignores all
identifiers, comments, and many keywords.
The neural networks used for code similarity analysis of other
benchmarks are similar to the one shown in
Figure~\ref{seq_tok_sim_fig}.  As we see in Table~\ref{table:seq_sim},
their accuracy is quite similar.

\textbf{SPT-based experiments}\\
\label{spt_experiments}
Following MISIM~\cite{misim}, the train, validation, and test datasets for the SPT-based experiments draw from entirely different problems.
In our experiments, we use 50\% problems for training, 25\% for validation, and 25\% for test. The train, validation, and test split used for the experiments can be found at~\cite{codenet}. Similarity scores in Table~\ref{table:code.similarity.2} and Table~\ref{table:code.similarity.3}
report mean and standard deviation of MAP@R~\cite{mapr} values evaluated with models trained using five random seeds. The models are trained on a Xeon(R) CPU E5-2680 v4, 2.4GHz, 256 GiB memory using a NVIDIA V100 GPU. The SPTs used in these experiments have nodes annotated with attributes derived by combining SPT features (refer to Section~\ref{sec-spt}), following the context-aware semantic structure~(CASS) proposed in~\cite{misim}. 

AROMA experiments are performed using the implementation of MISIM given in the further details section below ~\cite{gcj-297} and the input (SPTs) used for these experiments can be found at~\cite{codenet}. Due to the high memory requirement for computing MAP@R on the test set of CodeNet benchmarks, we had to reduce the feature set of AROMA. We estimate that AROMA results can improve by 10--25\% when all features are used. AROMA is rule-based and no training is involved, hence we don't report mean and standard deviation in Table~\ref{table:code.similarity.2}.  For each of the four datasets -- Java250, Python800, C++1000, C++1400 -- MISIM's GNN model is trained for a total of 1000 epochs at a learning rate of 0.001 with Adam optimizer. Each epoch consists of 1000 iterations, and in each iteration, 16 problems and 5 solutions per problem are randomly sampled, and all solution pairs are used for training as in ~\cite{misim}. MISIM results for the four languages can be reproduced by downloading the MISIM code and scripts~\cite{gcj-297} and using the provided CASS files~\cite{codenet} as input.  
    
For the GMN experiments (row 2 and row 3 in Table~\ref{table:code.similarity.3}), we adapt the implementation in~\cite{gmnimpl} of the GMN model~\cite{gmn} using SPTs~\cite{codenet} as graphs. We follow the recommendations in~\cite{gmn} for the model configuration, as they produce the best and stable results in our experiments. Specifically, we use 5 layers of propagation with weight sharing across layers, dot-product similarity for the cross-graph attention mechanism, and GRU layer to update node embeddings from the propagation scheme. For GMN training, given the large set of SPT pairs, we adopt an approach similar to~\cite{misim} of randomly sampling 16 problems with 5 solutions each. We use triplet loss with approximate hamming similarity~\cite{gmn} for each sample, which is formed using a similar pair combined with a dissimilar SPT. After every 100 iterations with a batch size of 64, another set of 16 problems and 5 solutions are sampled randomly for a total of 150,000 iterations (1500 sampled sets). GMN results could improve further with more training iterations. We use Adam optimizer with a learning rate of 1e-4 for training.

The first two rows of Table~~\ref{table:code.similarity.3} compare similarity models trained on SPT graph structure only. The first row in the table adapts the MISIM GNN model by masking the node labels to allow the model to learn structural features only. The second row uses the GMN~\cite{gmn} model with cross-graph attention-based matching for structural similarity using a node vector dimension of 32 and graph representation dimension of 128.

For the GMN+MISIM node attributes experiment, row 3 in Table~\ref{table:code.similarity.3}, we allow the GMN model to learn features based on both node attributes and the SPT structure. Accordingly, we replace the node encoder in the GMN, an MLP, with an embedding layer, for generating node feature vectors. We explore different node feature vector dimensions, such as 64, 100, 128, and found 100 to produce good results for the given number of training iterations. All other parameter settings remain the same as the structure only GMN experiments from row 2 of Table~\ref{table:code.similarity.3}. The GMN results can be reproduced using the Java250 CASS files available at~\cite{codenet}.

MAP@R score~\cite{mapr} is computationally expensive for GMN models because an embedding has to be computed for all SPT pairs in the test set, and hence Table~\ref{table:code.similarity.3} reports results on smaller sampled test sets.

\textbf{Details of MLM Experiment}\\
\label{appendix:mlm}
Here we show how a masked language model (MLM) can be trained with
CodeNet.  We closely follow the approach by Ankur Singh, documented in
the blog~\cite{singh}.  The goal of the model is to infer the correct
token for an arbitrary masked-out location in the source text.  We
assume that in every text, precisely one token is randomly masked. The
original token at such position is then the golden label.

From each of the 1000 C++1000 problems, we randomly select 100
samples for training and another 100 for testing.  Each C++ source
file is tokenized into a vocabulary of 442 distinct tokens as
categorized in Table~\ref{table:token_categories}.  For example,
\code{while} is a keyword and \code{strlen} is a function literal.

\begin{table}[h]
\small
\caption{Token categories used for MLM.}
\label{table:token_categories}
\centering
\begin{tabular}{|l|r|l|}
\hline
Type            & Count & Description \\
\hline
the keyword     &    95 & all C++20 reserved words \\
the function    &   280 & function names in common header files \\
the identifier  &    42 & standard identifiers, like stderr, etc. \\
the punctuator  &    16 & small set of punctuation symbols \\
\# or \#\#      &     2 & the C pre-processor symbols \\
0, 1            &     2 & special case for these frequent constants \\
the token class &     5 & identifier, number, operator, character, string \\
\hline
\end{tabular}
\end{table}

This code snippet:

\code{for (i = 0; i < strlen(s); i++) \{\}}

will be tokenized to:

\code{for ( id = 0 ; id < strlen ( id ) ; id operator ) \{ \}}

The tokenized source files are read into a pandas dataframe and
processed by the Keras Text Vectorization layer, to extract a
vocabulary and encode all token lines into vocabulary indices,
including the special ``[mask]'' token.  Each sample has a fixed token
length of 256. The average number of tokens per sample across the
training set is 474. Short samples are padded with 0 and those that
are too large are simply truncated.

The model is trained with 100,000 samples in batches of 32 over five
epochs, with a learning rate of 0.001 using the Adam optimizer.  We
evaluate the trained model on a test set of 100,000 samples.  Each
sample is pre-processed in the same way as the training samples and
one token (never a padding) is arbitrarily replaced by the ``[mask]''
symbol. Then, a prediction is generated and the top 1 and top 5
results are compared with the expected value. The achieved accuracies
are top-1: 0.9104 (stddev: 0.002) and top-5: 0.9935 (stddev: 0.0005).

\subsection{Generalization Across Datasets}
Models trained on the CodeNet benchmark datasets can benefit greatly from
their high quality. To demonstrate this, we compare
C++1000 to one of the largest publicly
available datasets of its kind, GCJ-297~\cite{gcj-297}. For the purpose of this comparison, we
train the same MISIM model on C++1000 and GCJ-297 and test the two
trained models on a third, independent dataset - POJ-104. The result
of this comparison is plotted in Figure~\ref{codenet_vs_gcj}.

The $x$-axis of this plot is the number of training epochs used and the
$y$-axis is the MAP@R score. The MISIM model for both datasets is
trained for $500$ epochs and the MAP@R score for validation and test
is computed after every ten epochs. There are a total of four
curves - a validation and a test curve for GCJ-297 and a validation
and a test curve for C++1000.

The training curves show that a $10$\% higher validation score can be
achieved with GCJ-297 compared to C++1000. However, when tested
on POJ-104, the model trained on GCJ-297 achieves a $12$\% lower score compared to the model trained on C++1000. We believe C++1000 has better generalization than GCJ-297 mainly for two
reasons: i) high data bias in GCJ-297 because the top $20$ problems with the most number of submissions account for $50$\% of all submissions and ii) cleaning and de-duplication of
submissions in CodeNet dataset (as described in Section~\ref{cleansing}).

\subsection{Masked Language Modelling for Token Inference}
A task such as code completion relies on the ability to predict a token at a
certain position in a sequence. To accomplish this we can build a masked
language model (MLM) using a technique that randomly masks out tokens in an
input sequence and aims to correctly predict them in an as-yet-unseen test set.
We train a popular BERT-like attention model on the C++1000 CodeNet benchmark after tokenization to a vocabulary of over 400 tokens and
obtain a top-1 prediction accuracy of 0.9104 (stddev: 0.002) and a top-5 accuracy of 0.9935 (stddev: 0.0005).







\section{Further Uses of CodeNet}
\label{use_cases}

The rich metadata and language diversity open CodeNet to a plethora of
use cases. The problem-submission relationship in CodeNet corresponds
to type-4 similarity~\cite{similarityType} and can be used for code
search and clone detection. The code samples in CodeNet are labeled
with their acceptance status so we can readily extract pairs of buggy and fixed code for code repair ~\cite{chen2018sequencer, yasunaga2021break}.  A large number of code samples come with inputs so that we can execute the code to extract the CPU run time and memory footprint, which can be used for regression studies and prediction.

CodeNet may also be used for program translation, given its wealth of programs written in a multitude of
languages. Translation between two programming languages is born out of a
practical need to port legacy codebases to modern languages in order to increase accessibility and lower maintenance costs. With the help of
neural networks, machine translation models developed for natural
languages~\cite{Wu2016} were adapted to programming languages,
producing pivotal success~\cite{Lachaux2020}. One considerable
challenge of neural machine translation is that model training depends
on large, parallel corpora that are expensive to
curate~\cite{Chen2018}, especially for low-resource languages (e.g.,
legacy code). Recently, monolingual
approaches~\cite{Lample2018,Lachaux2020} were developed to mitigate
the reliance on parallel data, paving ways to build models for
languages with little translation. Compared with 
current popular data sets (e.g., \cite{Lachaux2020,Lu2021}), CodeNet covers a much richer set of
languages with ample training instances.

\section{Conclusion}
Artificial intelligence has made great strides in understanding human language. Computer scientists have been fascinated by the possibility and tantalized by the vision of computers (AI) programming computers. In this paper, we presented "CodeNet", a first-of-its-kind very large-scale, diverse and high-quality dataset to accelerate the algorithmic advances in AI for Code. This dataset is not only unique in its scale, but also in the diversity of coding tasks it can help benchmark: from code similarity and classification for advances in code recommendation algorithms, and code translation between a large variety of programming languages, to advances in code performance improvement techniques. We hope that the scale, diversity and rich, high-quality annotations of CodeNet will offer unprecedented research opportunities at the intersection of AI and Software Engineering.

\section{Acknowledgements}
We would like to acknowledge AIZU and AtCoder for making the code submissions publicly available. We would like to thank the IBM Data Asset eXchange team for providing a platform to host the CodeNet dataset. We would like to thank the Women in Data Science team at Stanford University and the IBM Call for Code team for their collaboration in launching the CodeNet challenge.


\begin{thebibliography}{10}

\bibitem{allamanis2018survey}
Miltiadis Allamanis, Earl~T Barr, Premkumar Devanbu, and Charles Sutton.
\newblock A survey of machine learning for big code and naturalness.
\newblock {\em ACM Computing Surveys (CSUR)}, 51(4):1--37, 2018.

\bibitem{yang2020survey}
Yanming Yang, Xin Xia, David Lo, and John Grundy.
\newblock A survey on deep learning for software engineering.
\newblock {\em arXiv preprint arXiv:2011.14597}, 2020.

\bibitem{chen2021evaluating}
Mark Chen, Jerry Tworek, Heewoo Jun, Qiming Yuan, Henrique~Ponde
  de~Oliveira~Pinto, Jared Kaplan, Harri Edwards, Yuri Burda, Nicholas Joseph,
  Greg Brockman, Alex Ray, Raul Puri, Gretchen Krueger, Michael Petrov, Heidy
  Khlaaf, Girish Sastry, Pamela Mishkin, Brooke Chan, Scott Gray, Nick Ryder,
  Mikhail Pavlov, Alethea Power, Lukasz Kaiser, Mohammad Bavarian, Clemens
  Winter, Philippe Tillet, Felipe~Petroski Such, Dave Cummings, Matthias
  Plappert, Fotios Chantzis, Elizabeth Barnes, Ariel Herbert-Voss,
  William~Hebgen Guss, Alex Nichol, Alex Paino, Nikolas Tezak, Jie Tang, Igor
  Babuschkin, Suchir Balaji, Shantanu Jain, William Saunders, Christopher
  Hesse, Andrew~N. Carr, Jan Leike, Josh Achiam, Vedant Misra, Evan Morikawa,
  Alec Radford, Matthew Knight, Miles Brundage, Mira Murati, Katie Mayer, Peter
  Welinder, Bob McGrew, Dario Amodei, Sam McCandlish, Ilya Sutskever, and
  Wojciech Zaremba.
\newblock Evaluating large language models trained on code, 2021.

\bibitem{Lachaux2020}
Marie-Anne Lachaux, Baptiste Roziere, Lowik Chanussot, and Guillaume Lample.
\newblock Unsupervised translation of programming languages.
\newblock In {\em NeurIPS}, 2020.

\bibitem{wang2018machine}
Zheng Wang and Michael O’Boyle.
\newblock Machine learning in compiler optimization.
\newblock {\em Proceedings of the IEEE}, 106(11):1879--1901, 2018.

\bibitem{tier1contest}
\url{http://ibm.biz/cfcsc-codenet}.

\bibitem{wids}
Women in data science.
\newblock \url{https://widsconference.org/}.

\bibitem{aizu}
Yutaka Watanobe.
\newblock Aizu online judge.
\newblock \url{https://onlinejudge.u-aizu.ac.jp}.

\bibitem{atcoder}
Atcoder.
\newblock \url{https://atcoder.jp/}.

\bibitem{D2A}
Yunhui Zheng, Saurabh Pujar, Burn Lewis, Luca Buratti, Edward Epstein, Bo~Yang,
  Jim Laredo, Alessandro Morari, and Zhong Su.
\newblock D2a: A dataset built for ai-based vulnerability detection methods
  using differential analysis.
\newblock In {\em Proceedings of the ACM/IEEE 43rd International Conference on
  Software Engineering: Software Engineering in Practice}, ICSE-SEIP '21, New
  York, NY, USA, 2021. Association for Computing Machinery.

\bibitem{Devign}
Yaqin Zhou, Shangqing Liu, Jingkai Siow, Xiaoning Du, and Yang Liu.
\newblock Devign: Effective vulnerability identification by learning
  comprehensive programsemantics via graph neural networks.
\newblock In {\em Advances in Neural Information Processing Systems}, pages
  10197--10207. NeurIPS Foundation, 2019.

\bibitem{CTmaxmin}
Zhangyin Feng, Daya Guo, Duyu Tang, Nan Duan, Xiaocheng Feng, Ming Gonga,
  Linjun Shou, Bing Qin, Ting Liu, and Daxin Jiang.
\newblock Codebert: A pre-trained model for programming and natural languages.
\newblock {\em arXiv preprint arXiv:2002.08155v4}, 2020.

\bibitem{JavaCorpus}
Miltiadis Allamanis and Charles Sutton.
\newblock Mining source code repositories at massive scale using language
  modeling.
\newblock In {\em 10th Working Conference on Mining Software Repositories
  (MSR)}, page 207–216. IEEE, 2013.

\bibitem{Py150}
Veselin Raychev, Pavol Bielik, and Martin Vechev.
\newblock Probabilistic model for code with decision trees.
\newblock {\em ACM SIGPLAN Notices}, 2016.

\bibitem{Bugs2Fix}
Michele Tufano, Cody Watson, Gabriele Bavota, Massimiliano~Di Penta, Martin
  White, and Denys Poshyvanyk.
\newblock An empirical study on learning bug-fixing patches in the wild via
  neural machine translation.
\newblock In {\em ACM Transactions on Software Engineering and Methodology
  (TOSEM)}, pages 1--29, 2019.

\bibitem{CodeSearchNet}
Hamel Husain, Ho-Hsiang Wu, Tiferet Gazit, Miltiadis Allamanis, and Marc
  Brockschmidt.
\newblock Codesearchnet challenge: Evaluating the state of semantic code
  search.
\newblock {\em arXiv preprint arXiv:1909.09436v3}, 2019.

\bibitem{CONCODE}
Srinivasan Iyer, Ioannis Konstas, Alvin Cheung, and Luke Zettlemoyer.
\newblock Mapping language to code in programmatic context.
\newblock {\em arXiv preprint arXiv:1808.09588}, 2018.

\bibitem{codexglue}
Shuai Lu, Daya Guo, Shuo Ren, Junjie Huang, Alexey Svyatkovskiy, Ambrosio
  Blanco, Colin Clement, Dawn Drain, Daxin Jiang, Duyu Tang, Ge~Li, Lidong
  Zhou, Linjun Shou, Long Zhou, Michele Tufano, Ming Gong, Ming Zhou, Nan Duan,
  Neel Sundaresan, Shao~Kun Deng, Shengyu Fu, and Shujie Liu.
\newblock Codexglue: A machine learning benchmark dataset for code
  understanding and generation, 2021.

\bibitem{ncc}
Tal Ben-Nun, Alice~Shoshana Jakobovits, and Torsten Hoefler.
\newblock Neural code comprehension: A learnable representation of code
  semantics.
\newblock In S.~Bengio, H.~Wallach, H.~Larochelle, K.~Grauman, N.~Cesa-Bianchi,
  and R.~Garnett, editors, {\em Advances in Neural Information Processing
  Systems 31}, pages 3588--3600. Curran Associates, Inc., 2018.

\bibitem{gcj}
Farhan Ullah, Hamad Naeem, Sohail Jabbar, Shehzad Khalid, Muhammad~Ahsan Latif,
  Fadi Al-turjman, and Leonardo Mostarda.
\newblock Cyber security threats detection in internet of things using deep
  learning approach.
\newblock {\em IEEE Access}, 7:124379--124389, 2019.

\bibitem{misim}
Fangke Ye, Shengtian Zhou, Anand Venkat, Ryan Marcus, Nesime Tatbul,
  Jesmin~Jahan Tithi, Niranjan Hasabnis, Paul Petersen, Mattson. Timothy, Tim
  Kraska, Pradeep Dubey, Vivek Sarkar, and Justin Gottschlich.
\newblock Misim: A novel code similarity system, 2021.

\bibitem{poj-problem-samples}
\url{https://sites.google.com/site/treebasedcnn/home/problemdescription }.

\bibitem{gcj-297}
gcj-dataset.
\newblock
  \url{https://openreview.net/attachment?id=AZ4vmLoJft&name=supplementary_material}.

\bibitem{10.1145/3359591.3359735}
Miltiadis Allamanis.
\newblock The adverse effects of code duplication in machine learning models of
  code.
\newblock In {\em Proceedings of the 2019 ACM SIGPLAN International Symposium
  on New Ideas, New Paradigms, and Reflections on Programming and Software},
  Onward! 2019, page 143–153, New York, NY, USA, 2019. Association for
  Computing Machinery.

\bibitem{jaccard}
Wikipedia.
\newblock Jaccard index --- {W}ikipedia{,} the free encyclopedia.
\newblock \url{https://en.wikipedia.org/wiki/Jaccard_index}, 2020.

\bibitem{10.5555/2501720}
Terence Parr.
\newblock {\em The Definitive ANTLR 4 Reference}.
\newblock Pragmatic Bookshelf, 2nd edition, 2013.

\bibitem{aroma}
Sifei Luan, Di~Yang, Celeste Barnaby, Koushik Sen, and Satish Chandra.
\newblock Aroma: code recommendation via structural code search.
\newblock {\em Proceedings of the ACM on Programming Languages},
  3(OOPSLA):1–28, Oct 2019.

\bibitem{wala}
IBM T.J. Watson~Research Center.
\newblock Wala.
\newblock \url{https://github.com/wala/WALA}, 2021.

\bibitem{forbes}
Forbes on codenet.
\newblock
  \url{https://www.forbes.com/sites/moorinsights/2021/06/04/ibm-codenet-artificial-intelligence-that-can-program-computers-and-solve-a-100-billion-legacy-code-problem/?sh=343813636cdc}.

\bibitem{venturebeat}
Venturebeat on codenet.
\newblock
  \url{https://venturebeat.com/2021/05/10/ibms-codenet-dataset-aims-to-train-ai-to-tackle-programming-challenges/}.

\bibitem{zdnet}
Zdnet on codenet.
\newblock
  \url{https://www.zdnet.com/article/ibm-launches-autosql-watson-orchestrate-codenet-enterprise-ai-tools-at-think/}.

\bibitem{projectcodenet}
Project codenet repository.
\newblock \url{https://github.com/IBM/Project_CodeNet}.

\bibitem{cbert}
Luca Buratti, Saurabh Pujar, Mihaela Bornea, Scott McCarley, Yunhui Zheng,
  Gaetano Rossiello, Alessandro Morari, Jim Laredo, Veronika Thost, Yufan
  Zhuang, and Giacomo Domeniconi.
\newblock Exploring software naturalness through neural language models, 2020.

\bibitem{Kipf2017}
Thomas~N. Kipf and Max Welling.
\newblock Semi-supervised classification with graph convolutional networks.
\newblock In {\em ICLR}, 2017.

\bibitem{Xu2019}
Keyulu Xu, Weihua Hu, Jure Leskovec, and Stefanie Jegelka.
\newblock How powerful are graph neural networks?
\newblock In {\em ICLR}, 2019.

\bibitem{Thost2021}
Veronika Thost and Jie Chen.
\newblock Directed acyclic graph neural networks.
\newblock In {\em ICLR}, 2021.

\bibitem{DBLP:journals/corr/abs-2102-04664}
Shuai Lu, Daya Guo, Shuo Ren, Junjie Huang, Alexey Svyatkovskiy, Ambrosio
  Blanco, Colin~B. Clement, Dawn Drain, Daxin Jiang, Duyu Tang, Ge~Li, Lidong
  Zhou, Linjun Shou, Long Zhou, Michele Tufano, Ming Gong, Ming Zhou, Nan Duan,
  Neel Sundaresan, Shao~Kun Deng, Shengyu Fu, and Shujie Liu.
\newblock Codexglue: {A} machine learning benchmark dataset for code
  understanding and generation.
\newblock {\em CoRR}, abs/2102.04664, 2021.

\bibitem{li2018deeper}
Qimai Li, Zhichao Han, and Xiao-Ming Wu.
\newblock Deeper insights into graph convolutional networks for semi-supervised
  learning, 2018.

\bibitem{hu2020ogb}
Weihua Hu, Matthias Fey, Marinka Zitnik, Yuxiao Dong, Hongyu Ren, Bowen Liu,
  Michele Catasta, and Jure Leskovec.
\newblock Open graph benchmark: Datasets for machine learning on graphs.
\newblock {\em arXiv preprint arXiv:2005.00687}, 2020.

\bibitem{pyg}
Matthias Fey and Jan~E. Lenssen.
\newblock Fast graph representation learning with {PyTorch Geometric}.
\newblock In {\em ICLR Workshop on Representation Learning on Graphs and
  Manifolds}, 2019.

\bibitem{pytorch}
Adam Paszke, Sam Gross, Francisco Massa, Adam Lerer, James Bradbury, Gregory
  Chanan, Trevor Killeen, Zeming Lin, Natalia Gimelshein, Luca Antiga, Alban
  Desmaison, Andreas Kopf, Edward Yang, Zachary DeVito, Martin Raison, Alykhan
  Tejani, Sasank Chilamkurthy, Benoit Steiner, Lu~Fang, Junjie Bai, and Soumith
  Chintala.
\newblock Pytorch: An imperative style, high-performance deep learning library.
\newblock In H.~Wallach, H.~Larochelle, A.~Beygelzimer, F.~d\textquotesingle
  Alch\'{e}-Buc, E.~Fox, and R.~Garnett, editors, {\em Advances in Neural
  Information Processing Systems 32}, pages 8024--8035. Curran Associates,
  Inc., 2019.

\bibitem{xu2019powerful}
Keyulu Xu, Weihua Hu, Jure Leskovec, and Stefanie Jegelka.
\newblock How powerful are graph neural networks?, 2019.

\bibitem{similarityType}
Hitesh Sajnani.
\newblock {\em Large-Scale Code Clone Detection}.
\newblock PhD thesis, University of California, Irvine, 2016.

\bibitem{gmn}
Yujia Li, Chenjie Gu, Thomas Dullien, Oriol Vinyals, and Pushmeet Kohli.
\newblock Graph matching network for learning the similarity of graph
  structured objects.
\newblock In {\em International Conference on Machine Learning (ICML)}, 2019.

\bibitem{gmnimpl}
Graph-matching-networks.
\newblock \url{https://github.com/Lin-Yijie/Graph-Matching-Networks}.

\bibitem{mapr}
Kevin Musgrave, Serge~J. Belongie, and Ser{-}Nam Lim.
\newblock A metric learning reality check.
\newblock {\em CoRR}, abs/2003.08505, 2020.

\bibitem{codenet}
Codenet dataset.
\newblock \url{https://developer.ibm.com/exchanges/data/all/project-codenet}.

\bibitem{singh}
Ankur Singh.
\newblock "end-to-end masked language modeling with bert".
\newblock \url{https://keras.io/examples/nlp/masked_language_modeling}.

\bibitem{chen2018sequencer}
Zimin Chen, Steve Kommrusch, Michele Tufano, Louis-No{\"e}l Pouchet, Denys
  Poshyvanyk, and Martin Monperrus.
\newblock Sequencer: Sequence-to-sequence learning for end-to-end program
  repair.
\newblock {\em IEEE Transaction on Software Engineering}, 2019.

\bibitem{yasunaga2021break}
Michihiro Yasunaga and Percy Liang.
\newblock Break-it-fix-it: Unsupervised learning for program repair.
\newblock In {\em International Conference on Machine Learning (ICML)}, 2021.

\bibitem{Wu2016}
Yonghui Wu, Mike Schuster, Zhifeng Chen, Quoc~V. Le, Mohammad Norouzi, Wolfgang
  Macherey, Maxim Krikun, Yuan Cao, Qin Gao, Klaus Macherey, Jeff Klingner,
  Apurva Shah, Melvin Johnson, Xiaobing Liu, Łukasz Kaiser, Stephan Gouws,
  Yoshikiyo Kato, Taku Kudo, Hideto Kazawa, Keith Stevens, George Kurian,
  Nishant Patil, Wei Wang, Cliff Young, Jason Smith, Jason Riesa, Alex Rudnick,
  Oriol Vinyals, Greg Corrado, Macduff Hughes, and Jeffrey Dean.
\newblock Google's neural machine translation system: Bridging the gap between
  human and machine translation.
\newblock Preprint arXiv:1609.08144, 2016.

\bibitem{Chen2018}
Xinyun Chen, Chang Liu, and Dawn Song.
\newblock Tree-to-tree neural networks for program translation.
\newblock In {\em NeurIPS}, 2018.

\bibitem{Lample2018}
Guillaume Lample, Alexis Conneau, Ludovic Denoyer, and Marc'Aurelio Ranzato.
\newblock Unsupervised machine translation using monolingual corpora only.
\newblock In {\em ICLR}, 2018.

\bibitem{Lu2021}
Shuai Lu, Daya Guo, Shuo Ren, Junjie Huang, Alexey Svyatkovskiy, Ambrosio
  Blanco, Colin Clement, Dawn Drain, Daxin Jiang, Duyu Tang, Ge~Li, Lidong
  Zhou, Linjun Shou, Long Zhou, Michele Tufano, Ming Gong, Ming Zhou, Nan Duan,
  Neel Sundaresan, Shao~Kun Deng, Shengyu Fu, and Shujie Liu.
\newblock {CodeXGLUE}: A machine learning benchmark dataset for code
  understanding and generation.
\newblock Preprint arXiv:2102.04664, 2021.

\end{thebibliography}

\end{document}